\pdfoutput=1
\documentclass[a4paper,fleqn]{cas-sc}

\usepackage[numbers,sort&compress]{natbib}

\usepackage{amsmath,amssymb,amsfonts}
\usepackage{graphicx}
\usepackage{subcaption}
\usepackage{booktabs}
\usepackage{array}
\usepackage{multirow}
\usepackage{url}
\usepackage{xurl}

\begin{document}
\let\WriteBookmarks\relax
\def\floatpagepagefraction{1}
\def\textpagefraction{.001}

\shorttitle{GridForge: Constructing Open, Continental-Scale Synthetic Power Grids}
\shortauthors{J.~M.~Snodgrass et al.}

\title[mode=title]{GridForge: A Tractable Network-Flow Method\\ for Constructing
Open, Continental-Scale Synthetic Power Grids}

\author[1]{Jonathan M. Snodgrass}[orcid=0000-0003-2305-8136]
\cormark[1]
\ead{snodgrass@tamu.edu}
\credit{Conceptualization, Methodology, Software, Validation, Formal analysis,
Investigation, Data curation, Visualization, Writing -- original draft}

\author[1]{Thomas J. Overbye}[orcid=0000-0002-2382-2811]
\credit{Writing -- review and editing}

\author[2]{Christopher L. DeMarco}[orcid=0000-0002-8817-9804]
\credit{Conceptualization, Funding acquisition, Supervision, Writing -- review and
editing}

\author[2]{Bernard C. Lesieutre}[orcid=0000-0002-5274-324X]
\credit{Conceptualization, Funding acquisition, Supervision, Writing -- review and
editing}

\affiliation[1]{organization={Department of Electrical and Computer Engineering,
Texas A\&M University},
            city={College Station},
            state={TX},
            country={USA}}

\affiliation[2]{organization={Department of Electrical and Computer Engineering,
University of Wisconsin--Madison},
            city={Madison},
            state={WI},
            country={USA}}

\cortext[1]{Corresponding author.}

\begin{abstract}
The goal of this paper is to construct open, continental-scale synthetic
transmission grids with an optimization method that stays tractable. Security
restrictions keep models of the real grid out of open research and machine-learning
datasets, and optimization-based construction is hard at scale: the number of
candidate lines grows combinatorially, so electrically faithful methods prune the
candidates in advance. This paper presents GridForge, which keeps every bus pair
within physical length limits as a candidate, selects lines with Garver's
length-weighted network-flow linear program instead of binary decisions, and
resizes them with a sequential direct-current (DC) power flow. Two stages of 13
regional networks are evaluated. Their node-degree distributions are nearly as
strongly log-linear as those of the real Eastern and Western Interconnections
($R^2$ of $0.92$ to $0.98$ against $0.971$ and $0.981$), although the areas under
the fitted lines differ from the real ones by $39.9\%$ on average. On the upper
Midwest and Tennessee Valley footprints, resizing lowers the DC power-flow loss
from $5.89\%$ to $4.14\%$ and from $14.55\%$ to $7.05\%$, and the DC optimal
power-flow loss from $4.03\%$ to $2.02\%$ and from $5.54\%$ to $2.47\%$. Without
relaxing any line limit, the DC optimal power flow is feasible in $78.81\%$ and
$0.82\%$ of the scenarios. Most lines are built at 69~kV. The models span $250$ to
$78{,}484$ buses, were built on single machines, and are freely redistributable.
\end{abstract}

\begin{graphicalabstract}
\includegraphics[width=\linewidth]{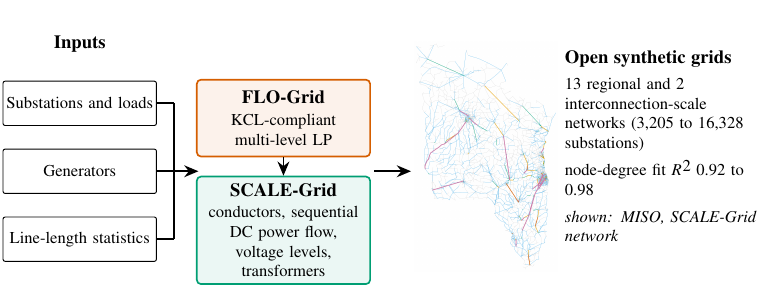}
\end{graphicalabstract}

\begin{highlights}
\item Garver's network-flow linear program picks lines among all length-feasible pairs
\item Each solution adds a forest of new lines; dispatch and outage scenarios add loops
\item Sequential DC power flow assigns multi-voltage conductor parameters
\item Resizing cuts the DC optimal power-flow loss from 4.0\% to 2.0\% and from 5.5\% to 2.5\%
\item Open, redistributable, continental-scale grids built on a single workstation
\end{highlights}

\begin{keywords}
Synthetic power grids \sep network construction \sep
network-flow linear programming \sep transportation model \sep
transmission expansion planning \sep data for machine learning
\end{keywords}

\maketitle

\section*{Nomenclature}

\noindent\textit{Sets and indices}\par\nopagebreak
\begin{tabular}{@{}l p{0.66\linewidth}@{}}
$n$              & Number of buses (vertices), indexed $\{1,\dots,n\}$ \\
$m$              & Number of existing (constructed) network elements (edges) \\
$G=(\mathcal{V},E)$ & Directed graph with $n$ vertices and $m$ edges \\
$\mathcal{V},E$  & Vertex (node) set and edge set of $G$ \\
$S,\,s$          & Set of net-injection scenarios; a scenario $s\in S$ \\
$_e,\,_p,\,_{p\|e}$ & Subscripts: existing, potential, and potential-parallel-to-existing paths \\
\end{tabular}

\medskip
\noindent\textit{Parameters}\par\nopagebreak
\begin{tabular}{@{}l p{0.66\linewidth}@{}}
$D,\,d$          & Bus-to-bus distance matrix / vector \\
$c,\,c_j$        & Vector of objective (cost) weights on branch flows; weight of branch $j$ \\
$p_j^{\max}$     & Thermal limit of branch $j$ \\
$x_j$            & Series reactance of branch $j$ (per unit) \\
$a_g,b_g,c_g$    & Constant, linear, and quadratic cost coefficients of generator $g$ \\
$C_{\mathrm{prod}},\,C_{\mathrm{cap}}$ & Production cost and capital cost of a network \\
$A_{\mathrm{e}},\,A_{\mathrm{p}}$ & Incidence matrices of the existing lines and of the candidate paths \\
$W,\,L,\,\alpha$ & Weight matrix and Laplacian of the Fiedler partition; weight of its distance term \\
$P$              & Vector of nodal net real-power injections, $P=P_G-P_D$ \\
$P_{D,s}(t)$     & Real-power load vector at time $t$, scenario $s$ \\
$P_{G,s}(t)$     & Generator real-power output at time $t$, scenario $s$ \\
$B_{\mathrm{Bus}}$ & Bus susceptance matrix \\
$A$              & Bus-branch incidence matrix \\
$G_{i,j},\,B_{i,j}$ & Real and imaginary parts of the bus admittance matrix entry $(i,j)$ \\
$\mathrm{GMD}$   & Geometric mean distance between phases \\
$\ell$           & Physical transmission-line length \\
\end{tabular}

\medskip
\noindent\textit{Variables}\par\nopagebreak
\begin{tabular}{@{}l p{0.66\linewidth}@{}}
$\zeta$          & Symmetric binary topology matrix; $\zeta_{i,j}=\zeta_{j,i}=1$ if an element joins buses $i$ and $j$ \\
$b$              & Branch (transmission-element) engineering-design parameters \\
$p$              & Vector of branch (element) power flows \\
$\theta$         & Vector of bus voltage phase angles \\
$|V_i|$          & Voltage magnitude at bus $i$ \\
\end{tabular}

\section{Introduction}\label{sec:intro}

After roughly two decades of essentially flat electricity demand, load in the
United States is now rising sharply, driven by artificial-intelligence and
hyperscale data centers, electrification, and the reshoring of manufacturing.
Global data-center electricity consumption is projected to more than double,
from about 415~TWh in 2024 to around 945~TWh by 2030 \cite{IEA_2025_EnergyAndAI},
and aggregate five-year US peak-load forecasts have risen roughly six-fold since
2022 \cite{Wilson_2025_NationalLoadGrowth}. Yet the interregional transmission
needed to serve this growth continues to be deployed slowly
\cite{Senga_2026_TransmissionExpansion}, and the median time from
interconnection request to commercial operation has grown to more than five
years \cite{Rand_2026_QueuedUp}.
Planning for this
growth will take many more studies, increasingly with machine-learning tools
trained on grid data, and both need large, realistic models of the electric power
system that can be shared freely.

For decades, the power-systems research community has relied on ``realistic but
not real'' synthetic mathematical network models to test new algorithms for a
wide range of power-system analyses
\cite{Birchfield_2023_RealisticGridSimulations,Taylor_2024_CATS}. Electric utility
companies, regional transmission operators (RTOs), and independent system
operators (ISOs) rely on such models of lines, transformers, generators, and loads
to operate and plan the system, since experiments on the actual power network are
neither desirable nor, in many cases, permitted. The restructuring of the sector
under the Energy Policy Act of 1992 and Federal Energy Regulatory Commission (FERC)
Orders 888 (1996) and 2000 (1999) separated vertically integrated utilities and
reduced information sharing. Planners now often expand transmission with
incomplete information about future generation and load growth.

Variable renewable generation adds to the need for tractable, large-scale
network-construction tools. The regions with the most favorable wind resource are
often not well connected electrically to the rest of the grid, and cost-optimized
continental siting studies find that higher levels of renewable energy require
substantial new long-haul transmission \cite{MacDonald_2016_FutureCostCompetitive}.
At the same time, the data needed to study such investments has become harder to
obtain. Models of the US transmission system were once publicly available and
could be freely shared. However, shortly after the September 11, 2001, terrorist
attacks, FERC established Critical Energy/Electric Infrastructure Information
(CEII) regulations that removed numerous documents describing the grid from public
circulation for the security of the nation's infrastructure \cite{FERC_CEII}. Restricted materials included FERC Form~715
\cite{FERC_Form715} and other databases correlating network parameters with
topological data. The concern is well founded: attackers mount more effective
attacks when both topological and component-parameter data are available than with
either alone \cite{Ouyang_2014_TopologyVulnerability}. Although researchers,
energy-market consultants, and other parties with a valid reason may request
CEII, the models cannot be published or released with any report or simulation,
so a model of the real network cannot serve as the shared basis for open
research. The community has therefore turned to synthetic grids. The legacy public
test cases are small and carry no geographic or component data, which limits their
use \cite{Babaeinejad_2021_PowerGridLibrary}, so researchers have sought large,
realistic, geolocated test systems instead.

The same restrictions apply to a fast-growing user of grid data: machine learning
for power systems. Recent benchmark datasets such as OPFData
\cite{Lovett_2024_OPFData} and PGLearn \cite{Klamkin_2025_PGLearn} build large
training sets by perturbing a small number of base cases. This varies the
operating point and the in-service topology, but every sample keeps the
construction of its base network. Both datasets are built for AC optimal power
flow, whereas the networks of this paper are sized against direct-current (DC)
power flow and need reactive planning, such as that of
\cite{Birchfield_2018_ReactivePowerPlanning}, before they can serve such datasets
(Section~\ref{sec:scale-transformers}). Proposed
``foundation models'' for the electric grid call for learning across diverse grid
data and topologies \cite{Hamann_2024_FoundationModelsGrid}, and the case for open
data and reproducible software in energy research \cite{Pfenninger_2017_OpenData}
points the same way. Every construction method leaves its mark on the grids it
builds: in the shape of the degree distribution, in the line-length and rating
mixes, and in how the voltage classes are laid out. A training set drawn from
several methods should therefore cover more of the space of plausible networks
than any one method can. That is a hypothesis, and testing it takes open models
from several construction methods, which means more methods able to build them.

Methods for constructing synthetic networks generally fall into two categories:
those that focus on the graph of the transmission network and those that focus on
its electrical performance. Graph-focused methods create large test systems at low
computational cost, and classical constructions such as small-world
\cite{Watts_1998_SmallWorld} and preferential-attachment
\cite{Barabasi_1999_ScalingRandomNetworks} graphs have been adapted to this end. However, these networks have proven less realistic than originally
anticipated. Comparisons of real grids against these abstract models show
substantial differences in degree distribution, clustering, and other properties,
so such models offer limited use for representing power grids
\cite{Hines_2010_TopologicalElectricalStructure,CotillaSanchez_2012_ComparingTopologicalElectrical}.
Methods that model electrical behavior yield networks whose realism can be checked
against the physics. The most developed of these build grids to match validation
metrics compiled from real systems
\cite{Birchfield_2017_GridStructuralCharacteristics,Birchfield_2017_MetricBasedValidation}
and have supplied the public case families on which much of the community now
depends \cite{Birchfield_2023_RealisticGridSimulations,TAMU_ElectricGridRepository}.
The construction problem carries a tension: algorithms that reproduce
graph statistics do not guarantee realistic electrical behavior, and formulations
that model electrical performance exactly rely on mixed-integer optimization that
is intractable at scale.

The key difficulty is combinatorial. For $n$ given
nodes, up to $\binom{n}{2}$ potential transmission branches could be added to
interconnect them. If every potential path is modeled, the
network admits up to
\begin{equation}\label{eq:combwall}
  2^{\binom{n}{2}}
\end{equation}
combinations of in-service and out-of-service branches, rendering the
construction problem intractable even for moderately sized systems. Existing
electrically faithful methods escape~\eqref{eq:combwall} by pruning: they limit
the potential connections in advance, cluster nodes to reduce $n$, or cap the size
of the network that can be modeled. In transmission expansion planning, an entire
body of work is devoted to reducing this combinatorial search space
\cite{Rahmani_2013_CombinatorialSearchSpace}, and the prevailing mixed-integer
formulations
\cite{Zhang_2012_MILPTransmissionExpansion,Bahiense_2001_MixedIntegerDisjunctive}
pre-select a modest candidate set to remain solvable. Unfortunately, pruning
decides before any optimization which paths a network may use, and a path that was
never considered cannot be built. What is needed is a computationally
tractable method that captures the electrical performance of the network
without a substantial a priori reduction in the number of candidate
branches.

\begin{figure}[t]
  \centering
  \includegraphics[width=\linewidth]{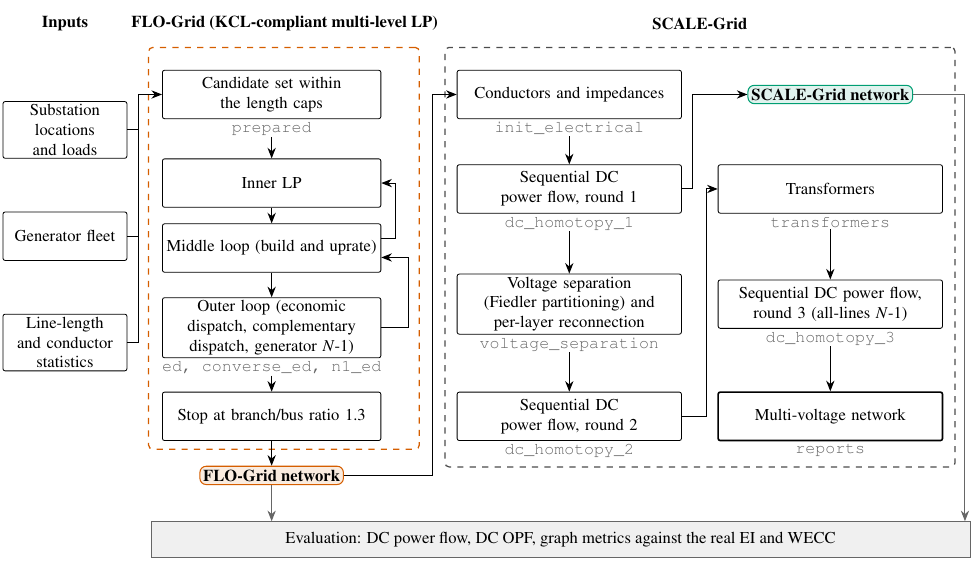}
  \caption{Overview of GridForge. FLO-Grid builds the topology and line limits with
  a KCL-compliant multi-level linear program over the candidate lines within the
  length caps, under economic dispatch, complementary dispatch, and generator $N$-1
  scenarios, and stops at a branch-to-bus ratio of 1.3 (the FLO-Grid network).
  SCALE-Grid assigns conductors and impedances, resizes lines with sequential DC
  power flow (the SCALE-Grid network after round 1), separates the voltage levels by
  Fiedler partitioning, reconnects each layer, and adds transformers. Gray type gives
  the GridForge stage names.}
  \label{fig-GridForge-Pipeline}
\end{figure}

This paper presents GridForge, a linear-programming-tractable methodology that
meets this need (Fig.~\ref{fig-GridForge-Pipeline}). Rather than assigning binary
in-service decisions to each candidate branch, GridForge keeps every bus pair
within the physical length limits as a candidate (Section~\ref{sec:data-regions}
lists the few other rules that remove candidates). It selects lines with the
classical network-flow, or transportation, model of transmission estimation
\cite{Garver_1970_TransmissionLP}: a linear program, here termed \emph{FLO-Grid},
that minimizes the length-weighted absolute branch flows (a weighted $L_1$ norm)
subject to Kirchhoff's current law (KCL). In a basic optimal solution of such a program, the candidate
lines that carry flow form a forest \cite{Ahuja_1993_NetworkFlows}, so a sparse set
of lines effectively ``switches on'' without any integer decisions. Repeated over
dispatch and generator-outage scenarios, these forests overlap into a meshed
network. A realism-and-scaling layer, termed \emph{SCALE-Grid}, then converts the
topology into a fully parameterized, multi-voltage network sized against DC power
flow.
SCALE-Grid assigns conductor parameters from statistical tables, tunes
transmission ratings through sequential DC power flow, creates voltage levels and
transformers, and assembles interconnection-scale models by spectral (Fiedler)
partitioning. Because its computational core is a linear program, GridForge
remains tractable on a single machine, without a compute cluster, across a wide
range of system sizes, handles both greenfield design and expansion, and yields
openly redistributable models that contain no CEII.

This paper develops in journal form the method first described in the first
author's doctoral dissertation \cite{Snodgrass_2021_Dissertation}: FLO-Grid, which
enforces KCL, and the SCALE-Grid parameterization layer, demonstrated on 13 regional
footprints. The complete open library of regional and interconnection-scale
models, with its full multi-metric validation, is left to future work
(Section~\ref{sec:conc}).

\subsection{Contributions}\label{sec:contributions}
The principal contributions of this paper are as follows.
\begin{enumerate}
  \item \textbf{No combinatorial candidate selection within a footprint.} Every
        bus pair within the per-voltage length caps is a candidate, apart from
        the rules listed in Section~\ref{sec:data-regions}. Lines are selected by
        Garver's network-flow linear program \cite{Garver_1970_TransmissionLP}
        rather than by binary decision variables, in contrast to the
        candidate-reduction strategies that classical mixed-integer expansion
        planning requires
        \cite{Rahmani_2013_CombinatorialSearchSpace,Zhang_2012_MILPTransmissionExpansion,Bahiense_2001_MixedIntegerDisjunctive}.
        Section~\ref{sec:flogrid} states what is inherited from
        \cite{Garver_1970_TransmissionLP,Villasana_1985_TransmissionLP} and what is new.
  \item \textbf{Tractable at regional scale.} The computational core stays a
        linear program, so footprints of up to $8441$ substations are built
        directly on a single machine and interconnection-scale models are
        assembled from them. SCALE-Grid then imposes electrical behavior through
        DC power flow instead of leaving it to graph statistics
        \cite{Hines_2010_TopologicalElectricalStructure,CotillaSanchez_2012_ComparingTopologicalElectrical}.
  \item \textbf{An end-to-end, multi-voltage pipeline.} GridForge proceeds from
        bare geolocated substations to a transformer-connected, multi-voltage
        interconnection designed to be $N$-1 feasible under DC power flow,
        including spectral-partition-based interconnection assembly.
        Section~\ref{sec:results} evaluates the first two stages of this
        pipeline (FLO-Grid and the first SCALE-Grid round); the later stages are
        described but not evaluated here.
  \item \textbf{Open, greenfield- and expansion-capable output.} The method
        generates unrestricted, redistributable models. Because its grids come
        from an optimization, not from statistical matching, recombination, or
        reconstruction, they differ from the open grids already in circulation
        and add to the range of networks available for benchmark and
        foundation-model work
        \cite{Babaeinejad_2021_PowerGridLibrary,Lovett_2024_OPFData,Klamkin_2025_PGLearn,Hamann_2024_FoundationModelsGrid}.
\end{enumerate}

The remainder of this paper is organized as follows.
Section~\ref{sec:bg} sets out the network-construction terminology, an idealized
optimization formulation of the construction problem, and the related work.
Section~\ref{sec:data} describes the input data and the statistics of real systems
that drive realism without redistributing any protected information.
Section~\ref{sec:flogrid} develops FLO-Grid, the KCL-compliant network-flow
linear program at the core of the method, with its multi-level architecture.
Section~\ref{sec:scalegrid} presents SCALE-Grid, the parameterization,
multi-voltage, and interconnection-assembly layer that turns a FLO-Grid topology
into a network sized against DC power flow.
Section~\ref{sec:results} demonstrates the method on 13 regional footprints and
compares its first two stages with the real Eastern and Western
Interconnections.
Section~\ref{sec:conc} concludes with a summary of findings, limitations, and
future work.

\section{Background and Idealized Network-Construction Framework}\label{sec:bg}
This section states network construction as an idealized optimization problem,
places GridForge on the hierarchy of power-flow relaxations, and states the gap.

\subsection{The four-step framework and an idealized formulation}\label{sec:bg-framework}
At the core of any network-construction algorithm (NCA) is the process of
assigning values to $\zeta$, the network topology variable. Independent of the
specific method, an NCA can be described as addressing four steps: (i) deciding
whether to place a new element on a given bus-to-bus path, i.e., assigning the
binary variable $\zeta_{i,j}$; (ii) assigning transmission-element parameters
such as per-unit impedance and thermal limit; (iii) determining the voltage
levels of the new elements and creating the buses within substations that are
joined by transformers; and (iv) assigning the physical construction parameters
of each element, such as conductor type and size and phase spacing. Depending on
the algorithm, these steps may be revisited iteratively. The complementary
problem of \emph{placing} the buses (creating or relocating the vertex set
$\mathcal{V}$ and assigning geographic coordinates) is treated here as an input
and is addressed separately (Section~\ref{sec:data}).

An idealized NCA would take the buses, the candidate-path lengths, and a set of
net-injection scenarios and solve, over a multi-year horizon, for the topology
$\zeta$, the element parameters $b$, and the generator dispatch $P_{G,s}(t)$
that minimize the combined production and capital cost subject to full
security-constrained AC power-flow feasibility, which includes Kirchhoff's voltage
law (KVL):
\begin{subequations}\label{eq-IdealNCA}
\begin{align}
\min_{\zeta,\,b,\,P_{G,s}(t)}\;&
  C_{\mathrm{prod}}\!\big(\zeta,b,P_{D,s}(t),P_{G,s}(t)\big)
  + C_{\mathrm{cap}}(\zeta,b) \label{eq-IdealNCA-obj}\\
\text{s.t.}\;& \zeta \in \{0,1\}^{n\times n},\ \ \zeta=\zeta^{\mathsf{T}}, \label{eq-IdealNCA-bin}\\
& N\!-\!1\ \text{security-constrained AC power-flow} \nonumber\\
& \text{feasibility (i.e., KCL, KVL and element} \nonumber\\
& \text{relations, together with line, voltage,} \nonumber\\
& \text{generator and stability limits) holding} \nonumber\\
& \text{for every contingency, scenario } s\in S \nonumber\\
& \text{and time } t\in T. \label{eq-IdealNCA-scpf}
\end{align}
\end{subequations}
The topology variable $\zeta$ is kept separate from the parameter variable $b$ so
that paths that are infeasible for real-world reasons (crossing large bodies of
water or environmentally sensitive terrain) can be excluded by fixing
$\zeta_{i,j}=0$, which limits the candidate paths without changing the nature of
the problem \cite{Tang_2015_HierarchicalFramework}. Tang and Ferris give a
multi-level hierarchical model that realizes many aspects of \eqref{eq-IdealNCA}
for transmission expansion as a mixed-integer nonlinear program (MINLP)
\cite{Tang_2015_HierarchicalFramework}; co-optimizing generation with
transmission, as integrated expansion planning does
\cite{Mahdavi_2023_IntegratedGT}, enlarges that formulation.

Formulation \eqref{eq-IdealNCA} is intractable for all but small systems, for two
reasons. First, retaining the full candidate graph admits up to
$2^{\binom{n}{2}}$ in-service/out-of-service configurations \eqref{eq:combwall}.
Second, the electrical constraints couple the binary and continuous variables:
modeling a branch flow through DC power flow as
$\zeta_{i,j}B_{i,j}(\theta_i-\theta_j)$ multiplies the binary decision
$\zeta_{i,j}$ by the continuous angles $\theta_i,\theta_j$, which yields a
mixed-integer program that is intractable for large $n$. Tractable NCAs
relax or reformulate the electrical constraints, the objective, or the
decision-variable domain \cite{Tang_2015_HierarchicalFramework}, and essentially all
electrically faithful methods prune the candidate set in advance.

\subsection{The relaxation ladder and DC power flow}\label{sec:bg-relaxation}
The tractability of an NCA is governed by how faithfully it models the AC
power-flow equations. The full AC relations at bus $i$ are
\begin{subequations}\label{eq-ACPF}
\begin{align}
P_i &= \textstyle\sum_{j=1}^{n} |V_i|\,|V_j|\,\big(G_{i,j}\cos(\theta_i-\theta_j)
      + B_{i,j}\sin(\theta_i-\theta_j)\big), \label{eq-ACPF-P}\\
Q_i &= \textstyle\sum_{j=1}^{n} |V_i|\,|V_j|\,\big(G_{i,j}\sin(\theta_i-\theta_j)
      - B_{i,j}\cos(\theta_i-\theta_j)\big), \label{eq-ACPF-Q}
\end{align}
\end{subequations}
where $P_i$ and $Q_i$ are the net real and reactive injections at bus $i$ and
$G_{i,j},B_{i,j}$ are the real and imaginary parts of the bus admittance matrix.
Progressively looser convex relaxations and linear approximations of
\eqref{eq-ACPF} form a well-studied hierarchy
\cite{Molzahn_2019_PowerFlowRelaxations,Coffrin_2016_NetworkFlowCopperPlate}:
from the full AC equations, to the second-order-cone (SOCP) relaxation, to the
linear DC power-flow approximation, to the network-flow (or transportation)
model that enforces only KCL and line limits, and
finally to the copper-plate model that enforces only aggregate power balance.
Each tighter rung gains fidelity at the cost of computational complexity.
The proposed method is positioned at two rungs of this ladder: its topology
optimization (FLO-Grid, Section~\ref{sec:flogrid}) sits at the network-flow rung,
while its parameterization layer (SCALE-Grid, Section~\ref{sec:scalegrid}) uses
DC power flow.

The DC power flow follows from \eqref{eq-ACPF-P} under the standard assumptions
that series resistance is negligible relative to reactance ($G\!\approx\!0$),
that angle differences are small so that $\cos(\theta_i-\theta_j)\approx1$ and
$\sin(\theta_i-\theta_j)\approx(\theta_i-\theta_j)$, and that per-unit voltage
magnitudes are near unity ($|V_i|\approx1$). Under these assumptions,
\eqref{eq-ACPF-Q} reduces to a constant and is neglected
\cite{Bergen_2000_PowerSystemsAnalysis,Coffrin_2016_NetworkFlowCopperPlate,Stott_2009_DCPowerFlowRevisited},
which leaves the linear relation
\begin{equation}\label{eq-DCPF}
  P_i = \sum_{\substack{j=1\\ j\neq i}}^{n} B_{i,j}\,(\theta_i-\theta_j).
\end{equation}

\subsection{Related work and the gap}\label{sec:bg-relatedwork}
Beyond the small-world \cite{Watts_1998_SmallWorld} and preferential-attachment
\cite{Barabasi_1999_ScalingRandomNetworks} constructions of Section~\ref{sec:intro},
graph-focused methods include statistically matched generators that target
realistic node-degree and line-impedance distributions
\cite{Wang_2010_StatisticallyCorrectTopologies}, an approach automated by the
AutoSynGrid toolkit for reference systems such as those of the New York ISO
(NYISO), the Western Electricity Coordinating Council (WECC), and the Electric
Reliability Council of Texas (ERCOT) \cite{Sadeghian_2020_AutoSynGrid}. Spatially embedded generators place synthetic
substations by a fitted spatial model and connect them by rules modeled on how
real grids grew \cite{Soltan_2015_SpatiallyEmbedded}, or grow a network by an
attachment rule that trades the cost of new lines against redundancy
\cite{Schultz_2014_RandomGrowth}. The Chung-Lu-Chain model reproduces prescribed
degree distributions and diameters across voltage levels
\cite{Aksoy_2019_GenerativeGraphModel} and is released in the open PowerGridSynth
package \cite{PowerGridSynth_2026}. A complex-network approach first builds a
low-cost network that meets demand and then reinforces it until its topology
matches that of real European grids \cite{Espejo_2019_ComplexNetwork}. Learned
generators sample topologies and parameters from a distribution fitted to real
grids, with a deep recurrent graph model \cite{Khodayar_2019_DeepGenerative} or,
more recently, with a graph diffusion process that reports power-flow convergence
for $98.9\%$ of its generated scenarios \cite{He_2025_PowerGrow}. Methods that do
not model electrical behavior scale readily but do not guarantee realistic power
flows
\cite{Hines_2010_TopologicalElectricalStructure,CotillaSanchez_2012_ComparingTopologicalElectrical}.

The most developed electrical-performance methods build grids to match
validation metrics compiled from real systems, an approach established by
Tom Overbye's group at Texas A\&M University
\cite{Birchfield_2017_GridStructuralCharacteristics,Birchfield_2017_MetricBasedValidation,Gegner_2016_GeographicallyRealistic}
and extended to many-voltage-level, multi-area systems
\cite{Birchfield_2018_BuildingManyVoltageLevels,Birchfield_2023_RealisticGridSimulations}.
These fit-to-footprint constructions set the realism standard for synthetic
grids and produced the public case families in widest use today
\cite{TAMU_ElectricGridRepository}. The line continues to develop: reactive
planning was added so that large synthetic grids converge under AC power
flow \cite{Birchfield_2018_ReactivePowerPlanning}, and planning sensitivities
were added to make the grids secure under contingencies and to target graph
properties directly
\cite{Birchfield_2020_PlanningSensitivities}.

Several other families of open synthetic grids are now in circulation, each built
by a different mechanism. Combined
transmission and distribution models extend the metric-matched approach down
through the distribution system \cite{Mateo_2024_TexasSyntheticTD}. Geographically
anchored constructions place invented topology and parameters onto the real
corridors of a region, so that the grid is geo-referenced while corresponding to
no actual line \cite{Taylor_2024_CATS}. Recombination methods assemble synthetic
networks from anonymized fragments of real utility models
\cite{PNNL_2018_SDET}. Reconstruction pipelines recover a network from public map
data and calibrate it until it solves, yielding models that trace the observed
grid closely \cite{Britto_2026_GridSFM}. Anonymized real-derived cases remain in
circulation as a fifth kind \cite{Josz_2016_PEGASE}. These families differ in what
they keep from real grids and what they invent, and a benchmark library that draws
on several of them covers ground that none covers alone
\cite{Babaeinejad_2021_PowerGridLibrary}.

However, the optimization-based literature is dominated by the need to prune.
The network-flow (transportation) model traces to Garver's linear program
\cite{Garver_1970_TransmissionLP}, later hybridized with DC power flow on the
existing lines \cite{Villasana_1985_TransmissionLP}. Subsequent
transmission-expansion-planning (TEP) formulations that add DC power-flow
constraints resort to sensitivity-based heuristics to select a small candidate
set \cite{Romero_2005_ConstructiveHeuristicDC}, and mixed-integer formulations
\cite{Zhang_2012_MILPTransmissionExpansion,Bahiense_2001_MixedIntegerDisjunctive}
likewise pre-select a modest set of paths
\cite{Rahmani_2013_CombinatorialSearchSpace}. In industry practice, the candidate
set is a short list from the start. In the planning processes of ERCOT, the
Midcontinent Independent System Operator (MISO), and the Southwest Power Pool
(SPP), transmission owners and other stakeholders propose specific projects for
each identified need, and the planner evaluates those proposals and their
alternatives \cite{ERCOT_2025_PlanningProcess,MISO_2026_MTEP,SPP_2026_DPP}.
Neither these formulations nor GridForge
(Section~\ref{sec:flogrid-complexity}) guarantees a globally optimal network; they
differ in where the candidate set is cut.

The open synthetic grids reviewed above are built by statistical matching,
metric-matched construction, recombination, and reconstruction. To the authors'
knowledge, none is produced by solving a network-construction optimization over the
candidate graph of a footprint; the cost-aware growth and reinforcement rules of
\cite{Schultz_2014_RandomGrowth,Espejo_2019_ComplexNetwork} come closest. The reason
is tractability: formulated exactly, that optimization is the intractable program
\eqref{eq-IdealNCA}, and the expansion-planning literature that comes closest to
solving it stays tractable by pruning the candidate set in advance. Instead,
GridForge applies Garver's network-flow linear program to every length-feasible bus
pair of a footprint (Section~\ref{sec:flogrid}) and recovers multi-voltage
electrical realism through a DC-power-flow parameterization layer
(Section~\ref{sec:scalegrid}), whose inputs are described next. Note that this
paper does not claim its grids are more realistic than those already available,
and Section~\ref{sec:results} shows where they are not. The claim is that a
tractable construction optimization produced them, and that their differences from
other families trace back to its objective.

\section{Input Data and Statistical Characterization}\label{sec:data}
This section describes the data GridForge uses and how the statistics that drive
its realism are compiled from public and protected sources. Note that \emph{no CEII
is ever redistributed}: the method extracts only aggregate statistics
(distributions of line lengths, MVA ratings, and conductor parameters) from the
protected data, and these statistics carry no topological or geographically
identifiable information.

\subsection{Required inputs}\label{sec:data-inputs}
GridForge takes two classes of input and produces a transmission-network graph
with design parameters $b$. At the substation level, it needs the substations to
be connected, the length of every potential path between them (from the
substations' geographic coordinates), and a set of net-injection scenarios that
combine generator cost curves and power limits with time-varying load. At the
branch level, it needs a ``lookup'' table of candidate conductors, each with
per-unit-length impedance values and an MVA limit. The bus-to-bus distance of each
candidate path stands in for the cost of new transmission
(Section~\ref{sec:flogrid-cost}). The set of existing lines may be empty, in which
case a greenfield, fully synthetic network is constructed.

\subsection{Substation and load planning}\label{sec:data-substation}
The substation locations and the time-varying load that seed the construction
are themselves synthesized upstream and are treated here as given inputs.
Synthetic substations come from US Census Bureau tract data, and hourly active and
reactive load is disaggregated to each bus from state-level public load data
\cite{Snodgrass_2026_SyntheticLoad,Li_2021_LoadTimeSeries} (Supplementary
Section~S1.4).

\subsection{Real-world data sources}\label{sec:data-sources}
Branch-level input statistics and the metrics used to validate realism come from
three sources that describe the real North American Eastern and Western
Interconnections. The public FERC Form~1 \cite{FERC_Form1} reports per-line
lengths, voltage levels, and conductors, but no impedances and no topology.
FERC Form~715 \cite{FERC_Form715}, classified as CEII \cite{FERC_CEII}, gives
line MVA ratings, transformer impedances, and line lengths inferred from the
line impedances, but no geography. The U.S. Department of Energy (DOE) state energy-sector
risk profiles \cite{DOE_2014_StateRiskProfiles} cross-check the Form~1 mileage.
Supplementary Section~S1.5 tabulates what each source
supplies and derives the Form~715 line lengths.

\subsection{Synthetic regions and maximum line lengths}\label{sec:data-regions}
Because constructing a model of a complete interconnection is intractable
without a large compute cluster, 13 geographic footprints were defined to
resemble the territories operated by present-day ISOs, RTOs, and balancing
authorities (Table~\ref{tab:substations_by_region}). Each footprint contains
roughly 3000 to 9000 synthetic substations and is drawn to be
geographically ``convex'', so that few candidate paths cross large bodies of
water or states with no intermediate substations. The two to three most common
high-voltage levels ($\ge\!230$~kV) of each region come from the Form~1 and DOE
data. Supplementary Section~S1.1 maps every footprint.

\begin{table}[t]
\centering
\caption{Synthetic test systems: the 13 regional footprints and the two
combined systems. Footprint names are those of the U.S. Environmental Protection
Agency's eGRID subregions (AZNM, FRCC, NWPP, RFCE, RFCW, SRMV, SRTV, SRVC) and of
the ISOs and RTOs (CAISO, MISO, NEISO, NYISO, SPP) whose territories they
resemble. Load is the mean over the $|S|$ evaluation scenarios of the
FLO-Grid network. CUSA (SRTV, MISO, and SRMV) is not evaluated electrically. The
combined systems have more buses than substations because their voltage levels
are separated and joined by transformers.}
\label{tab:substations_by_region}
\small
\begin{tabular}{@{}l>{\raggedright\arraybackslash}p{4.4cm}rrrrrr@{}}
\toprule
\textbf{Region} & \textbf{States} & \textbf{Substations} & \textbf{Buses} & \textbf{Branches} & \textbf{Branch/bus} & \textbf{Load (GW)} & $|S|$ \\
\midrule
AZNM & Arizona, Colorado, Nevada, New Mexico, Utah & 4{,}719 & 4{,}719 & 5{,}971 & 1.265 & 27.8 & 486 \\
CAISO & California & 8{,}404 & 8{,}404 & 10{,}558 & 1.256 & 29.2 & 486 \\
FRCC & Florida, Georgia, South Carolina & 7{,}440 & 7{,}440 & 9{,}674 & 1.300 & 45.8 & 486 \\
MISO & Illinois, Iowa, Minnesota, Wisconsin & 6{,}968 & 6{,}968 & 9{,}051 & 1.299 & 36.0 & 486 \\
NEISO & Connecticut, Maine, Massachusetts, New Hampshire, Rhode Island, Vermont & 3{,}501 & 3{,}501 & 4{,}203 & 1.201 & 15.9 & 486 \\
NWPP & Idaho, Montana, Oregon, Washington, Wyoming & 3{,}205 & 3{,}205 & 3{,}960 & 1.236 & 25.8 & 86 \\
NYISO & New York & 5{,}008 & 5{,}008 & 6{,}142 & 1.226 & 20.4 & 485 \\
RFCE & New Jersey, Pennsylvania & 5{,}371 & 5{,}371 & 6{,}967 & 1.297 & 31.3 & 486 \\
RFCW & Indiana, Michigan, Ohio & 7{,}409 & 7{,}409 & 9{,}632 & 1.300 & 47.8 & 486 \\
SPP & Kansas, Nebraska, North Dakota, Oklahoma, South Dakota, Texas & 8{,}441 & 8{,}441 & 10{,}974 & 1.300 & 68.1 & 486 \\
SRMV & Arkansas, Louisiana, Missouri & 3{,}291 & 3{,}291 & 4{,}279 & 1.300 & 24.7 & 486 \\
SRTV & Alabama, Kentucky, Mississippi, Tennessee & 4{,}625 & 4{,}625 & 6{,}015 & 1.301 & 34.7 & 486 \\
SRVC & DC, Delaware, Maryland, North Carolina, Virginia, West Virginia & 6{,}520 & 6{,}520 & 8{,}477 & 1.300 & 45.1 & 486 \\
\midrule
CUSA & SRTV + MISO + SRMV & 14{,}884 & 21{,}367 & 36{,}644 & 1.715 & --- & --- \\
WECC & NWPP + AZNM + CAISO & 16{,}328 & 20{,}757 & 32{,}711 & 1.576 & 91.5 & 115 \\
\bottomrule
\end{tabular}
\end{table}

Each voltage class has a length beyond which construction is impractical, because
the angular separation across a line is bounded by stability limits (typically
$30^{\circ}$ to $45^{\circ}$) and its reactance grows with length
\cite{StClair_1953_LineLoadability,Dunlop_1979_Loadability}.
Table~\ref{table-Rules} lists the maximum length adopted for each voltage level,
with the other rules and parameters of the code that built the evaluated
networks.
The maxima were chosen mostly from the 98th to 99th percentile of the Form~715
inferred lengths, with the lower voltages set nearer the 90th to 97th percentile of
the Form~1 lengths (Supplementary Sections~S1.3 and~S1.7). The longest of
them pre-zeroes the entries of $\zeta$ for longer pairs, and each level's maximum
later sets the lowest voltage a line of a given length may use
(Section~\ref{sec:data-conductor}). Apart from these length limits, three
rules remove candidates (Table~\ref{table-Rules}). Pairs of generating
substations more than $80$~mi apart are removed, and shorter ones are weighted
$\times 2.5$ (Section~\ref{sec:flogrid-incidence}). Candidate paths parallel to
lines at other voltage levels are removed when a voltage level is reconnected
(Section~\ref{sec:scale-reconnect}). Systems larger than about $8500$ substations
are partitioned into footprints, so no candidate joins two footprints
(Section~\ref{sec:scale-interconnect}). Each rule is physical or practical; none
selects candidates combinatorially within a footprint. At a $230$~mi cap, the
candidate set keeps $24\%$ (AZNM) to $93\%$ (NEISO) of all bus pairs, from 2.2
million candidates in SRMV to 20.1 million in RFCW (Supplementary Section~S1.7 and
Table~S12).

\begin{table}[tp]
\centering
\caption{Rules and parameters of FLO-Grid and SCALE-Grid in the code that built
the evaluated networks. Lengths in miles are converted to kilometers with the
factor 1.61; $d$ is the haversine distance between two buses.}
\label{table-Rules}
\small
\begin{tabular}{@{}>{\raggedright}p{4.0cm}>{\raggedright\arraybackslash}p{11.9cm}@{}}
\toprule
\textbf{Rule} & \textbf{Value} \\
\midrule
\multicolumn{2}{@{}l}{\emph{Inputs and candidate set}} \\
Candidate set & Every bus pair; haversine distance, R = 6,371 km \\
Length cap; per-kV maxima & d < max(MaxLength)$\cdot$1.61 km; MaxLength 30 / 50 / 90 / 130 / 180 / 230 mi for 69 \ldots{} 765 kV \\
Gen--gen length cap & Gen--gen pairs with d > 80$\cdot$1.61 km removed \\
Gen--gen penalty & Short ($\leq$ 80 mi) gen--gen weights $\times$ 2.5 \\
Path costs & New 5d (NewLineCost = 5), built corridor 3d ($\times$ 0.6), built line 1d \\
\midrule
\multicolumn{2}{@{}l}{\emph{FLO-Grid inner LP and middle loop}} \\
Inner LP & min $\sum_j c_j(p_j^{+}+p_j^{-})$ s.t. $A_{\mathrm{e}}(p_{\mathrm{e}}^{+}-p_{\mathrm{e}}^{-})+A_{\mathrm{p}}(p_{\mathrm{p}}^{+}-p_{\mathrm{p}}^{-})=P$; $0\le p_{\mathrm{e}}^{\pm}\le p^{\max}$ on existing lines; candidate paths capped at $10^{6}$ MVA (unbounded in effect) \\
Lines per pass & Candidates sorted by abs(flow); each is kept unless its From or To bus already received a new line in the pass \\
New-line rating & max(ceil(1.5$\cdot$flow), NewLineCapacity) \\
Upgrade of an existing line & Candidate parallel to an existing line: limit += ceil(1.25$\cdot$flow) \\
Middle-loop stop & No new line, and the largest selected abs(flow) < 15 MVA or fewer than $5k$ candidate paths carry flow ($k$: lines picked in the pass) \\
\midrule
\multicolumn{2}{@{}l}{\emph{FLO-Grid outer loop}} \\
Phases run & Economic dispatch (ED) $\rightarrow$ complementary dispatch (CED) $\rightarrow$ generator N-1 under ED $\rightarrow$ generator N-1 under CED \\
Complementary threshold & Units with (Pmax -- x)/Pmax > 0.15 get lb = 0.99$\cdot$Pmax; others lb = 0 \\
Generator N-1 outages & Every generator bus, ascending bus index \\
Target branch/bus ratio & MinBusBranch = 1/1.3, checked after each outage and inside the loop \\
\midrule
\multicolumn{2}{@{}l}{\emph{Conductors and impedances}} \\
Conductor choice & Voltage floor from length (> MaxLength(kV)$\cdot$1.61 forces the next level), then the closest MVA \\
GMD draw & GMDIndex = randi([1, 1000]) per line; column floor(kV/115)+1 (765 kV $\rightarrow$ 6) \\
R/X/B model & z = Ra + j(Xa + Xd(GMD)); y = j/((Xa$^{\prime}$ + Xd$^{\prime}$(GMD))$\cdot$1e6); Z = zL$\cdot$sinh($\gamma$L)/($\gamma$L); B = Im(yL$\cdot$tanh($\gamma$L/2)/($\gamma$L/2)); base kV$^2$/100 \\
\midrule
\multicolumn{2}{@{}l}{\emph{SCALE-Grid (sequential DC power flow, round 1)}} \\
DC power flow & B-$\theta$, bus 1 as slack; flows = 100$\cdot$B\_prim$\cdot$A$^{\mathsf{T}}$$\theta$ (MW, 100 MVA base) \\
Upsize batch & ceil(min(max(2 \%$\cdot$n\_over, 10), n\_over)); earlier copies 1.5 \% and 0.5 \% \\
Downsize batch & ceil(min(max(1.5 \%$\cdot$n\_under, 5), n\_under)); earlier copies 1.0 \% and 0.5 \% \\
Angle limit and angle pass & MaxAngleDC1 = 35$^\circ$; over-angle lines get +1 conductor per phase, re-snapped and stability-locked \\
L$\infty$ envelope & Elementwise max over the 7 capacity sets, re-snap by length, R/X/B \\
\midrule
\multicolumn{2}{@{}l}{\emph{Voltage separation}} \\
Fiedler Laplacian & $L = \mathrm{diag}(W\mathbf{1}) - W$, $W = \alpha/d$ (all pairs, km) + line MVA; $\alpha = 10$ \\
\midrule
\multicolumn{2}{@{}l}{\emph{Transformers}} \\
Bus split & One bus per voltage level; the highest voltage keeps the substation bus; lower-voltage lines move to new buses \\
\bottomrule
\end{tabular}

\smallskip
\begin{minipage}{\linewidth}\footnotesize
Values are from the archived 2019 code closest in date to the evaluated networks. The
candidate length cap and the per-voltage length table varied by region
(Section~\ref{sec:data-regions}), and the round-1 batch fractions varied between $0.5\%$ and
$2\%$. Voltage separation and transformers are not part of the FLO-Grid and
SCALE-Grid networks evaluated in Section~\ref{sec:results}; their rows are from the later code.
\end{minipage}
\end{table}

\subsection{Conductor lookup tables and impedance}\label{sec:data-conductor}
For each region, two conductor lookup tables were assembled from the public
data: one sets the voltage level at each substation, and one, with a wider range
of ratings (including multi-conductor bundles), resizes conductors within a
voltage class. The MVA limit of each candidate line is approximated from the
Form~1 conductor size and type using manufacturer ampacity data
\cite{Southwire_2012_ACSR}, and its per-unit-length reactance components are
taken from a transmission-line reference book
\cite{LaForest_1982_TransmissionLineReferenceBook}. The phase-spacing term of each
line is drawn from a mileage-weighted distribution of the Form~1 structure data, so
every constructed line receives realistic impedance and rating values without
reproducing any protected model (Supplementary Section~S1.6). A physically
consistent reconstruction of conductor and structure parameters for synthetic lines
is developed further in \cite{Diop_2026_TransmissionLineDesign}.

\section{FLO-Grid: KCL-Compliant Network-Flow Linear Program}\label{sec:flogrid}

FLO-Grid, the first of the method's two stages, is a tractable relaxation of the
idealized construction problem~\eqref{eq-IdealNCA} that keeps only KCL: the net
real power entering each bus must sum to zero. It sits on the
network-flow (transportation) rung of the relaxation ladder of
Section~\ref{sec:bg-relaxation}. FLO-Grid decouples the production-cost term
of~\eqref{eq-IdealNCA-obj} and evaluates it separately with a generator economic
dispatch (Section~\ref{sec:flogrid-outer}), which leaves an objective that depends
only on the constructed lines $\zeta$ and their design parameters $b$. Following
Garver \cite{Garver_1970_TransmissionLP}, the binary in-service decisions on $\zeta$
are replaced with a linear program (LP) over branch flows. Here the LP covers every
bus pair of a footprint within the physical length limits
(Section~\ref{sec:data-regions}), without the a priori candidate pruning that
mixed-integer expansion planning requires.

Table~\ref{table-GarverDelta} separates what FLO-Grid inherits from the
transportation model of \cite{Garver_1970_TransmissionLP} and its hybrid with DC
power flow \cite{Villasana_1985_TransmissionLP} from what this paper adds. Garver's
procedure solves the LP over the existing lines and a set of overload paths, assigns
a new circuit to the path with the largest flow, and repeats until no overload path
carries flow \cite{Garver_1970_TransmissionLP}. FLO-Grid keeps the LP and the
add-and-repeat loop and changes the candidate set, the number of lines added per
solution, the cost classes, and the injection scenarios; the parameterization and
assembly stages of SCALE-Grid that follow are not part of
\cite{Garver_1970_TransmissionLP} or \cite{Villasana_1985_TransmissionLP}.

\begin{table}[t]
\centering
\caption{What FLO-Grid inherits from the transportation model and what this paper
adds.}
\label{table-GarverDelta}
\small
\begin{tabular}{@{}>{\raggedright}p{3.0cm}>{\raggedright}p{5.6cm}>{\raggedright\arraybackslash}p{7.0cm}@{}}
\toprule
\textbf{Element} & \textbf{Garver \cite{Garver_1970_TransmissionLP}, Villasana et al. \cite{Villasana_1985_TransmissionLP}} & \textbf{FLO-Grid and SCALE-Grid} \\
\midrule
Inner problem & LP: minimize length-weighted absolute flows subject to KCL and existing-line limits & Same LP \eqref{eq-LNCA-MinL1} \\
Solution method & Bus guide potentials over a spanning tree & Standard-form LP, general LP solver (CPLEX) \\
Candidate set & Overload paths chosen by the planner & Every bus pair within the length caps of a footprint of up to $8441$ substations (Section~\ref{sec:data-regions}) \\
Cost classes & Two: existing lines and overload paths & Three: $1d$ existing, $3d$ parallel to existing, $5d$ new (Section~\ref{sec:flogrid-cost}) \\
Lines added per solution & One, on the path with the largest flow & Several, in order of decreasing flow, at most one new line per substation (Section~\ref{sec:flogrid-middle}) \\
Injection scenarios & One net-injection vector & Economic, complementary, and generator $N$-1 dispatches, to a branch-to-bus ratio of 1.3 (Section~\ref{sec:flogrid-outer}) \\
DC power flow & On the existing lines, in the hybrid model \cite{Villasana_1985_TransmissionLP} & After construction, to resize every line (Section~\ref{sec:scale-sdcpf1}) \\
Parameters, voltage levels, transformers, assembly & Not part of the method & SCALE-Grid (Section~\ref{sec:scalegrid}) \\
\bottomrule
\end{tabular}
\end{table}

\subsection{Multi-level optimization architecture}\label{sec:flogrid-arch}
\begin{figure}[t]
  \centering
  \includegraphics{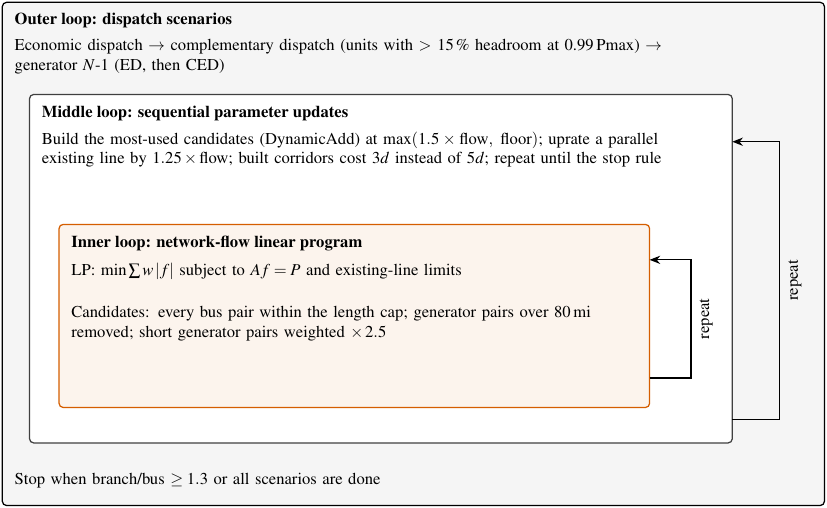}
  \caption{The three nested loops of FLO-Grid: the inner network-flow linear
  program, the middle loop that builds the most-used candidate lines and uprates
  existing ones, and the outer loop over the dispatch scenarios and generator $N$-1
  contingencies, which stops at a branch-to-bus ratio of 1.3.}
  \label{fig-FLOGrid-Loops}
\end{figure}

The geolocated buses and their data (Section~\ref{sec:data}) are passed to a
multi-level algorithm that fixes the topology by assigning $\zeta$, comprising
three nested loops (Fig.~\ref{fig-FLOGrid-Loops}). The \emph{inner
loop} solves the linear program~\eqref{eq-LNCA-MinL1} for the power flows $p$ on
the existing lines and the candidate paths. The \emph{middle loop} uses those
flows to decide which paths to add and which existing lines to upgrade, and
assigns their capacity limits. When no further lines are added or upgraded, the
\emph{outer loop} updates the net-injection vector across generator-dispatch
scenarios and generator $N$-1 contingencies. The algorithm terminates when a
target branch-to-bus ratio is met or all scenarios are exhausted. The
parameterization stages that follow (Fig.~\ref{fig-GridForge-Pipeline}) form the
SCALE-Grid layer of Section~\ref{sec:scalegrid}; they are heuristic and do not change
the optimization formulation presented here.

\subsection{Inner loop: the network-flow linear program}\label{sec:flogrid-inner}
The inner loop solves the optimization problem~\eqref{eq-LNCA-MinL1}, which takes
the nodal net-injection vector $P$ (Section~\ref{sec:flogrid-outer}), the objective
weights $c$ (Section~\ref{sec:flogrid-cost}), and the incidence matrix $A$
(Section~\ref{sec:flogrid-incidence}):
\begin{subequations}\label{eq-LNCA-MinL1}
\begin{align}
\min_{p}\quad & \textstyle\sum_{j} c_j \lvert p_j\rvert \label{eq-ObjFunL1}\\
\text{s.t.}\quad & A\,p = P, \label{eq-KCLConstr}\\
& \lvert p_j\rvert \le p_j^{\max}\ \ \text{for each existing line } j,
   \label{eq-LineLim}
\end{align}
\end{subequations}
where $p$ is the vector of branch flows and $p_j^{\max}$ the thermal limit of existing
line $j$. Constraint~\eqref{eq-KCLConstr}
enforces KCL at every bus. The thermal limit~\eqref{eq-LineLim} applies
\emph{only} to existing lines, leaving candidate-path flows unconstrained so the LP
is free to route power onto any corridor. The injections must also balance,
$\sum_{i=1}^{n} P_i = 0$ (the copper-plate condition of
Section~\ref{sec:bg-relaxation}); this is a condition on the data, met because $P$
is formed from a peak-load scenario and an economic dispatch.

The objective~\eqref{eq-ObjFunL1} is a rough surrogate for the combined capital and
operational costs of~\eqref{eq-IdealNCA-obj}: the unit weights on existing lines
approximate loss-driven operational cost, and the larger weights on candidate paths
approximate new-construction capital cost (Section~\ref{sec:flogrid-cost}); it is
the MW-mile objective of the transportation model \cite{Garver_1970_TransmissionLP}.
Its sparsity follows from the structure of a minimum-cost network-flow problem. Such a
linear program has a basic optimal solution, the kind a simplex solver returns, in
which the branches whose flows lie strictly between their bounds form a forest (a
set of trees with no loops) \cite{Ahuja_1993_NetworkFlows}. Candidate paths have no
flow limit, so the candidate paths that carry flow in that solution form a forest:
at most $n-1$ of the $\binom{n}{2}$ candidates, with zero flow on every other one.
A sparse subset of candidate lines is therefore selected for construction
\emph{without any binary decision variables}, and no candidate path within a
footprint is removed beyond the rules of Section~\ref{sec:data-regions}. The same
property makes each solution radial; the loops of the final network come from
repeating the program over the scenarios of the outer loop
(Section~\ref{sec:flogrid-outer}).

\subsection{Incidence-matrix construction}\label{sec:flogrid-incidence}
Two incidence matrices represent the flows on existing lines and on candidate
paths, $A_{\mathrm{e}}$ for the existing lines and $A_{\mathrm{p}}$ for the
complete graph of candidate paths \cite{Garver_1970_TransmissionLP}, concatenated
column-wise; an empty $A_{\mathrm{e}}$ denotes a greenfield, fully synthetic network.
An existing line and its parallel candidate path produce identical columns, so the
thermal limit~\eqref{eq-LineLim} binds the existing line while the parallel path
remains free to carry a second circuit. The candidate-removal rules of
Section~\ref{sec:data-regions} are applied when $A_{\mathrm{p}}$ is built
(Supplementary Section~S3.4).

\subsection{Objective-function weighting factors}\label{sec:flogrid-cost}
Because the capital cost of a new line is approximately proportional to its length,
FLO-Grid uses the haversine bus-to-bus distance $d$ as a surrogate for construction cost
\cite{Garver_1970_TransmissionLP,McCalley_2014_TransmissionTechnologies}. The
weight vector partitions the branches into three classes,
\begin{equation}\label{eq-CostVector}
  c =
  \begin{bmatrix}
    d_{\mathrm{e}} \\ 3\,d_{\mathrm{p}\parallel\mathrm{e}} \\ 5\,d_{\mathrm{p}}
  \end{bmatrix},
\end{equation}
where $d_{\mathrm{e}}$, $d_{\mathrm{p}\parallel\mathrm{e}}$, and $d_{\mathrm{p}}$
are the lengths of existing lines, of candidate paths parallel to an existing line,
and of candidate paths with no parallel existing line, respectively. The weight $1d$
on existing lines is a rough surrogate for losses, $5d$ approximates the per-mile
cost of new construction, and $3d$ the lower cost of upgrading a corridor. Values from
$2d$ to $5d$ were tested on two to three test cases, and $3d$ best reproduced the
target node-degree distribution. It also matches the upper range of
\cite{Hanson_1991_UpgradingTransmission} and the lower range of
\cite{SCE_2014_UnitCostGuide} (Supplementary Section~S3.1). The node-degree
comparison of Section~\ref{sec:results-nodedegree} is therefore not an independent
test of this weight. Two further quantities are calibration targets: the
branch-to-bus ratio of 1.3 at which the outer loop stops
(Section~\ref{sec:flogrid-complexity}), which sets the mean node degree, and the
high-voltage line length and $69$~kV line share that steer the SCALE-Grid resizing
(Section~\ref{sec:scale-sdcpf1}). The DC power-flow and DCOPF results of
Section~\ref{sec:results-electrical} over the 484 hourly scenarios were not used to
tune the method.

\subsection{Middle loop: sequential parameter updates}\label{sec:flogrid-middle}
The middle loop uses the flows $p$ returned by the inner loop to add lines and to
update the thermal limits~\eqref{eq-LineLim} and the cost weights $c$, after which
control returns to the inner loop. In a modification of the procedure of
\cite{Garver_1970_TransmissionLP}, the middle loop adds the candidate paths in
order of decreasing flow, at most one new line per substation in each pass. Each
new line receives a thermal limit of $150\%$ of its path flow or the smallest rating in the lookup
table, whichever is larger. Paths parallel to a newly added line drop from $5d$ to
$3d$, and an existing line whose parallel path carries flow is uprated by $125\%$
of that flow. The loop stops when no line is added and either the largest
selected flow is below $15$~MVA or few candidate paths still carry flow
(Table~\ref{table-Rules}; Supplementary Section~S3.5). On its own, this inner--middle formulation yields
networks more radial than real grids; the outer loop supplies the missing loops.

\subsection{Outer loop: generation and load scenarios}\label{sec:flogrid-outer}
The outer loop computes and updates the net-injection vector $P$, taking loads as
negative injections and generators as positive. A forecast peak-load condition
drives the scenarios. Each state's annual peak load is allocated to its load buses
by census-tract share and raised by a planning margin of up to $25\%$, which gives
a non-coincident peak above the hourly peak of the companion load data
\cite{Snodgrass_2026_SyntheticLoad} (Supplementary Sections~S1.2 and~S3.6).
The first scenario pairs this peak with a standard generator economic dispatch
\cite{Bergen_2000_PowerSystemsAnalysis}. A second, \emph{complementary}, dispatch
then holds every generator whose economic-dispatch output is below $85\%$ of its
maximum at $99\%$ of that maximum, and sets the lower bound of every other
generator to zero, so that the network can deliver power from any generator.
Further iterations take each generating substation out in turn, in bus order,
re-dispatch, and run the inner and middle loops to convergence. This sequence runs first
under the economic and then under the complementary dispatch, progressively
meshing the network (Supplementary Sections~S3.6 and~S3.3).

\subsection{Termination and computational considerations}\label{sec:flogrid-complexity}
The multi-level algorithm terminates when the target branch-to-bus ratio is reached
or all generator-dispatch and $N$-1 scenarios have run. Because the computational
core is an LP rather than a mixed-integer program, individual regions of up to about
8500 substations are constructed directly on a single machine, without a compute cluster, and
interconnection-scale models are assembled from such regions
(Section~\ref{sec:scale-interconnect}). Because only KCL and thermal limits are
enforced, the scheme is heuristic and carries no global-optimality guarantee, and its
transportation-model flows can differ substantially from those of DC or AC power
flow \cite{Neumann_2022_PowerFlowLossApproximations}.
The SCALE-Grid layer of Section~\ref{sec:scalegrid} assigns physical parameters and
tunes ratings to remove most DC-power-flow overloads.

\section{SCALE-Grid: Parameterization, Multi-Voltage, and Interconnection Assembly}\label{sec:scalegrid}

SCALE-Grid converts a FLO-Grid topology (the in-service decisions $\zeta$ and the
branch-flow limits of Section~\ref{sec:flogrid}) into a complete multi-voltage
network with impedances, buses, transformers, and line ratings sized against a DC
power flow (the SCALE-Grid box of Fig.~\ref{fig-GridForge-Pipeline}).
It carries out steps (ii) to (iv) of the four-step framework of
Section~\ref{sec:bg-framework}, which FLO-Grid deferred.

In this layer, the voltage levels are \emph{not} fixed in advance at each
substation. Instead, the DC power flow on each line selects its conductor from the
lookup tables of Section~\ref{sec:data-conductor}, and the selected conductor
carries a voltage level. A line's voltage class can rise or fall as the
sequential DC power flow tunes its rating (Section~\ref{sec:scale-sdcpf1}). Because
the lookup table spans all voltage levels present, the transmission backbone and
its voltage structure develop together from the DC power flow. Two calibration
targets steer the resizing (Section~\ref{sec:scale-sdcpf1}), so the voltage mix is
in part a calibration result.

\subsection{Branch parameterization}\label{sec:scale-param}
Once the topology is fixed, the candidate paths are removed from the
linear-programming model, and the MVA limit of each remaining line selects the
closest conductor in the lookup table of Section~\ref{sec:data-conductor}. The impedance
$(R,X,B)$ follows from that conductor, the straight-line length $\ell$ between the
two substations, and a phase spacing (GMD) drawn from the per-region Form~1
distribution \cite{FERC_Form1,LaForest_1982_TransmissionLineReferenceBook}.
Every constructed element then carries an underlying voltage level, conductor
size (kcmil), phase spacing, and length (miles), together with impedance $(R,X,B)$
and a thermal limit (Supplementary Section~S4.4).

\subsection{Sequential DC power flow: tuning transmission}\label{sec:scale-sdcpf1}
This step updates the branch parameters $(R,X,B)$ and the thermal limits so that few
or no branches are overloaded across the net-injection scenarios. SCALE-Grid solves
a DC power flow~\eqref{eq-DCPF} and compares each line's flow with its limit. After
each solution, it upgrades a subset of the overloaded lines (the
larger of $2\%$ of them or $10$ lines) to the next-higher-rated conductor. It
downgrades a subset of the underused lines (the larger of $1.5\%$ of them or five
lines) to the next-smaller conductor. A line counts as underused when its flow is
below $25\%$ to $40\%$ of its rating, a margin set by how far the high-voltage line
length falls short of its target (Supplementary Section~S4.5 and Table~S3).
The lines to resize are chosen \emph{at random}, which gives a distinct parameter set
on each run and realistic rating distributions (Supplementary Section~S4.5); the
2019 runs set no random seed, so the run-to-run spread of the evaluated networks
cannot be reported (Section~\ref{sec:conc}). Two
targets steer the resizing: the share of $69$~kV lines is kept between $58\%$ and
$68\%$, and the resizing stops only once the total high-voltage (230 to 765~kV) line
length reaches a target taken from the FERC Form~1 and DOE data (Supplementary
Table~S3). Iteration then continues until at most $0.5\%$ of the lines and
no more than 25 lines are overloaded and fewer than $10\%$ are underused.
The seven sub-runs of the dispatch, angle, and $N$-1 passes are combined per branch
by the maximum rating (the $L_\infty$ norm), fixing the final $(R,X,B)$
(Supplementary Section~S4.5).
The resizing can also be run under a transmission-line $N$-1 analysis, which is
designed to make the network DC $N$-1 feasible; Section~\ref{sec:results} does not
test this property. Overloaded lines that already carry the highest-rated conductor
of their class are flagged for manual double-circuiting (Supplementary
Section~S4.5). By adding
branch parameters and enforcing a power-flow solution, this step introduces
Kirchhoff's voltage law and the network element relations, placing the model on
the DC-power-flow rung of the relaxation ladder (Section~\ref{sec:bg-relaxation}).

\subsection{Creating buses with assigned voltage levels}\label{sec:scale-buses}
Because impedances are absent from the FLO-Grid linear program, voltage levels are
assigned only during the sequential DC power flow, and to the \emph{conductors}
rather than the buses. A substation therefore acquires a separate bus for each
distinct voltage class among its incident lines, initially joined by ideal
transformers (Section~\ref{sec:scale-transformers}). Because this assignment ignores
connectivity, the sub-network at a given voltage level is frequently \emph{not}
connected, as in the $138$~kV network of the Wisconsin model (Supplementary
Section~S4.2), and must be reconnected before the model is usable.

\subsection{Reconnection by re-running FLO-Grid per voltage level}\label{sec:scale-reconnect}
To reconnect a voltage level, its sub-network is returned to the FLO-Grid
multi-level algorithm (Section~\ref{sec:flogrid}) to add lines. An exact reduction
such as the Kron reduction \cite{Bergen_2000_PowerSystemsAnalysis} would route no
power onto candidate paths. Instead, each bus at the level being reconnected gets a
random \emph{proxy} injection of the same order of magnitude as the summed DC flows
on its substation's lines at all other levels. Candidate paths
parallel to lines at other levels are removed first, so that voltage levels do not
stack along one corridor. The sequential DC power flow is then re-run with each line
held within its own voltage class, and residual overloads are corrected manually or
by the line $N$-1 pass (Supplementary Section~S4.5). With the PowerWorld edits to
the delivered Eastern case (Section~\ref{sec:results-nodedegree}), these are the
manual steps of the method.

\subsection{Partitioning voltage levels by the Fiedler eigenvector}\label{sec:scale-fiedler}
For larger systems, reconnecting an entire voltage level as one graph can require
lines longer than the physical maxima of Table~\ref{table-Rules}, so each level is
first partitioned by a Fiedler-eigenvector method. A symmetric, weighted Laplacian
$L$ is formed from the line capacities and the bus-to-bus distances,
\begin{equation}\label{eq-Laplacian}
\begin{aligned}
  &L = \mathrm{diag}(W\mathbf{1}) - W, \qquad W = W_d + W_\ell,\\[0.5ex]
  &W_{d,i,j}=
  \begin{cases}
    \alpha / D_{i,j}, & i \neq j,\\
    0, & i = j,
  \end{cases}\\[0.5ex]
  &W_{\ell,i,j}=
  \begin{cases}
    \text{MVA limit of the element joining buses } i,j, & \text{if it exists},\\
    0, & \text{otherwise}\ (i = j\ \text{included}),
  \end{cases}
\end{aligned}
\end{equation}
with $\alpha$ a positive constant, $\mathbf{1}$ the vector of ones, and $D$ the full,
symmetric bus-to-bus distance matrix.
The distance term evens the partition sizes and guarantees that $L$ has rank
$n-1$. Buses are split by the sign of the Fiedler vector, the eigenvector of the
second-smallest eigenvalue of $L$ \cite{Fiedler_1973_AlgebraicConnectivity}, which
induces a bipartition that is connected in the weighted graph of $W$
\cite{Fiedler_1975_EigenvectorsGraphTheory} (the basis of spectral bisection
\cite{Pothen_1990_PartitioningSparse} and spectral clustering
\cite{vonLuxburg_2007_SpectralClustering}); the procedure recurses on each
partition. Because $W_d$ joins every pair of buses, the guarantee does not extend
to the lines, and a partition may hold several groups of buses joined by lines,
which the reconnection step then joins (Supplementary Section~S4.8). A value of
$\alpha = 10$ minimizes the tie lines between partitions while keeping the
partitions similar in size.

\subsection{Adding transformers and a note on AC feasibility}\label{sec:scale-transformers}
After the second sequential DC power flow, impedances and MVA limits are assigned to
the ideal transformers at each substation. The rating comes from the sums of the
ratings of the lines, generators, and transformers on each side, and the per-unit
impedances and $X/R$ ratios come from the FERC Form~715 statistics
\cite{FERC_Form715}. The released Wisconsin $1664$-bus system was produced by this
complete pipeline (Supplementary Section~S4.9). At this point, the network carries
the parameters and injections required for an AC power-flow solution and is
designed to have a feasible DC optimal power flow (DCOPF); Section~\ref{sec:results}
evaluates the earlier FLO-Grid and SCALE-Grid stages, not this one. Full AC
power-flow or AC optimal power-flow feasibility requires reactive-power support such
as switched shunt capacitors or static VAr compensators, which lies outside the
scope of the present method; the reactive-planning procedure of
\cite{Birchfield_2018_ReactivePowerPlanning} can be applied.

\subsection{Interconnection-scale assembly}\label{sec:scale-interconnect}
\begin{figure}[t]
  \centering
  \includegraphics[width=0.62\linewidth]{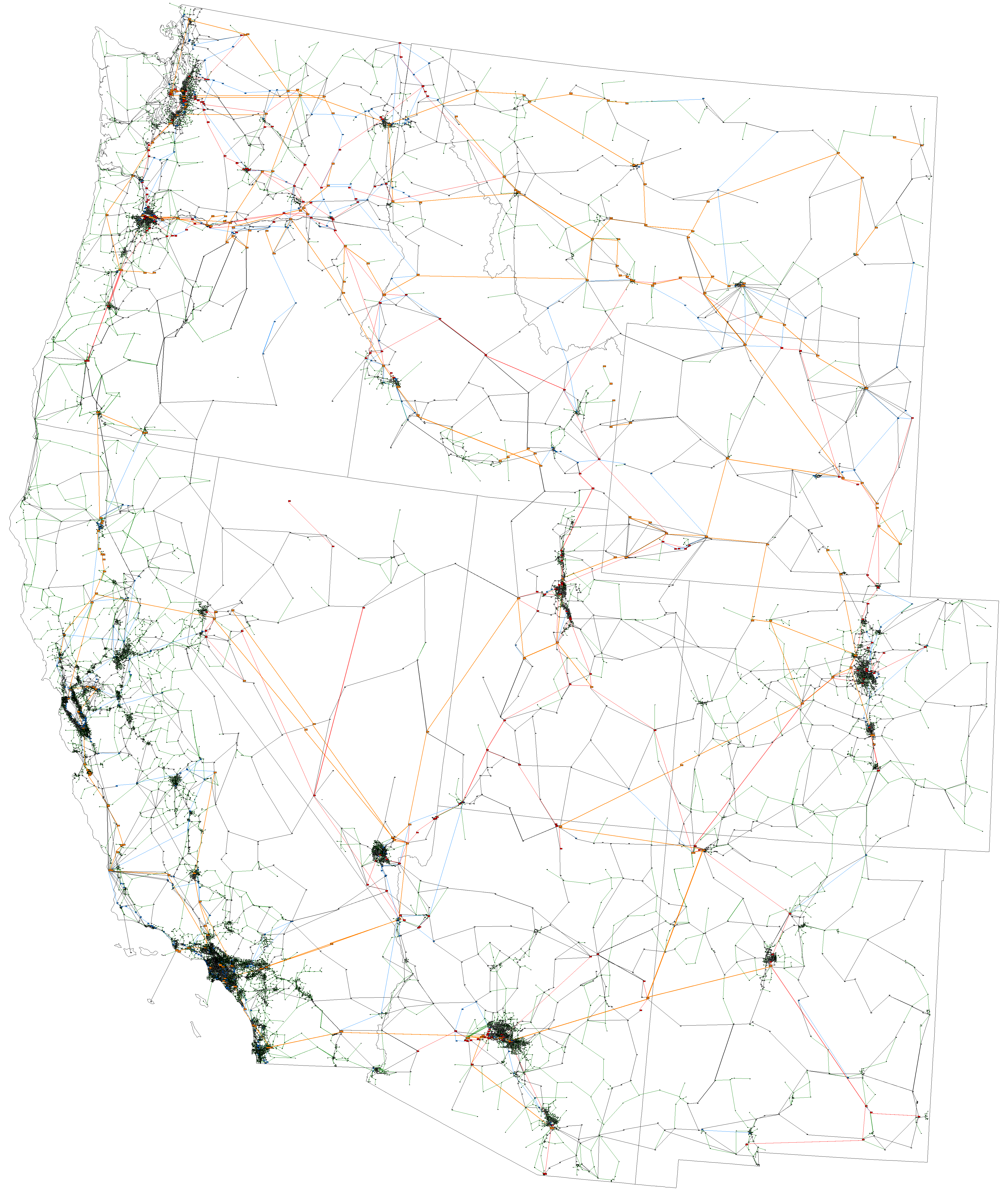}
  \caption{One-line diagram of the delivered $20{,}758$-bus synthetic Western United
  States case after Fiedler partitioning and interconnection of the three regional
  areas, colored by voltage level (black: 69~kV; dark green: 138~kV; blue: 230~kV;
  red: 345~kV; orange: 500~kV; gray: state borders).}
  \label{fig-WECC-OneLine}
\end{figure}

Systems larger than about $8500$ substations are assembled from regional pieces.
The FLO-Grid multi-level algorithm and the first sequential DC power flow are run
on each footprint to establish its topology and its buses with voltage levels. The
voltage sub-networks of the separate
footprints are then combined and reconnected, an approach also pursued in
\cite{Birchfield_2018_BuildingManyVoltageLevels} and mirroring the historical
development of transmission systems as interconnections of smaller regional
networks. Each voltage level is partitioned by the Fiedler method of
Section~\ref{sec:scale-fiedler} until its sub-networks can be joined by lines shorter
than the maximum length for that class (Table~\ref{table-Rules}). The remaining
reconnection steps yield a network designed to be $N$-1 feasible under DC power
flow, with a feasible DCOPF, properties not evaluated in Section~\ref{sec:results}. For the synthetic
Western United States model, three regional areas were built first (California,
the Pacific Northwest, and the Desert Southwest) and then interconnected
(Supplementary Section~S4.3). The result is the $20{,}757$-bus WECC system
evaluated in Section~\ref{sec:results}; the delivered $20{,}758$-bus Western case
of Fig.~\ref{fig-WECC-OneLine} is the same network after PowerWorld edits that
added four midpoint buses and removed three. Its $500$~kV network is a single
connected network, and its $230$ and $345$~kV networks are partitioned. Given
sufficient computing time, the initial partitioning would be unnecessary; it is a
hardware-practicality decision rather than a fundamental limit of the method.
Networks built in this way range from a $250$-bus New England system to a
$78{,}484$-bus Eastern United States network; Supplementary Table~S11 lists every
network, its stage, and whether it is evaluated in Section~\ref{sec:results} or
released.

\section{Case Study and Demonstration}\label{sec:results}

This section demonstrates GridForge and measures the realism of its output against
real networks. Synthetic models were constructed for the 13 geographic footprints of
Section~\ref{sec:data-regions} (Table~\ref{tab:substations_by_region}) and, by
combining adjacent footprints through the assembly of
Section~\ref{sec:scale-interconnect}, for three larger systems approximating the
Eastern Interconnection, the Western Interconnection, and the central United
States. Footprints are made of whole states, so SPP includes all of Texas, ERCOT
included (5492 of its 8441 substations). The delivered Eastern case therefore
contains a Texas system as one area tied to the rest by 69 AC branches, whereas
the real ERCOT system is a separate interconnection.
Each synthetic substation sits at the centroid of a US census tract, so substation
density, and with it network quality, varies by footprint; the six footprints shown
as one-line diagrams were chosen to span that range, and statistics are given for
all 13. Throughout, the benchmarks of realism are the real
Eastern and Western Interconnection network models assembled from the FERC
Form~715 data of Section~\ref{sec:data-sources}~\cite{FERC_Form715}, following the
metric-based validation philosophy established for synthetic
grids~\cite{Birchfield_2017_MetricBasedValidation,Birchfield_2017_GridStructuralCharacteristics}.

\subsection{Experimental setup}\label{sec:results-setup}
The evaluation covers two stages of every regional network, not the delivered
multi-voltage cases of Section~\ref{sec:scale-interconnect}. The first, the
\emph{FLO-Grid} network, is the topology produced when the multi-level
network-flow linear program of Section~\ref{sec:flogrid} terminates, with branch
impedances assigned by the initial conductor lookup of
Section~\ref{sec:scale-param}. The second, the \emph{SCALE-Grid} network, is the
same topology after the first round of sequential DC power flow in SCALE-Grid
(Section~\ref{sec:scale-sdcpf1}) has resized conductors and ratings against the
DC power flow, before the voltage levels are
separated.
Reporting both stages isolates the contribution of the parameterization layer.
Every metric is reported on true line lengths, the straight-line distance
between substations.

The net-injection scenarios that drive construction are the economic- and
complementary-dispatch cases and the generator $N$-1 contingencies of
Section~\ref{sec:flogrid-outer}. Each footprint was evaluated on 486 scenarios
(NYISO on 485 and NWPP on 86; Supplementary Section~S5.5): the economic- and
complementary-dispatch design scenarios and 484 hourly operating scenarios. The
hourly scenarios are taken from the synthetic hourly load around the annual peak
and minimum and committed by a simple unit commitment with a $10\%$ spinning
reserve.
Every reported value is the mean over the footprint's scenarios. Because the DC
power flow treats branches as lossless, the evaluation recovers approximate
real-power losses after the fact by multiplying the square of each branch current
by its resistance and converting to MW. Each inner-loop program was solved with
CPLEX~\cite{IBM_2017_CPLEX} (versions 12.6.3, 12.7, and 12.9, as recorded in the
run logs) called from MATLAB, on
single machines ranging from lab desktops to 32-core servers with 256~GB of
memory. The DC power-flow and DCOPF evaluations follow the standard
formulations~\cite{Zimmerman_2011_MATPOWER}.

For MISO and SRTV, the FLO-Grid and SCALE-Grid metrics are exact: the impedances
of both networks were recomputed with true line lengths and evaluated on all 486
scenarios. For the other 11 footprints, they are rescaled estimates, marked
$^{\dagger}$ in the tables and by hollow markers in the figures, because the 2021
evaluations of these footprints used other lengths. The FLO-Grid impedances applied
the kilometer-to-mile conversion twice, so their lengths were 1.61 times too short,
and the SCALE-Grid networks were built and evaluated with lengths 1.4 times the
straight-line distance, an allowance for routing. The hourly injections of these 11
footprints were not kept, so their networks cannot be re-evaluated. Instead, each
scenario's loss is rescaled to true line lengths (by 1.61 and by 1/1.4), and the
flow-based metrics keep their 2021 values. A uniform scaling of the branch
impedances would leave the DC flows unchanged; the long-line model is not exactly
linear in length, so on MISO and SRTV the flow-based metrics move by at most
$0.3\%$.
To check the estimate, it was also computed for MISO and SRTV, where it lies
within $0.026$ percentage points of the exact loss (Supplementary Section~S5.5).

\subsection{Qualitative realism}\label{sec:results-oneline}
\begin{figure}[tbp]
  \centering
  \includegraphics{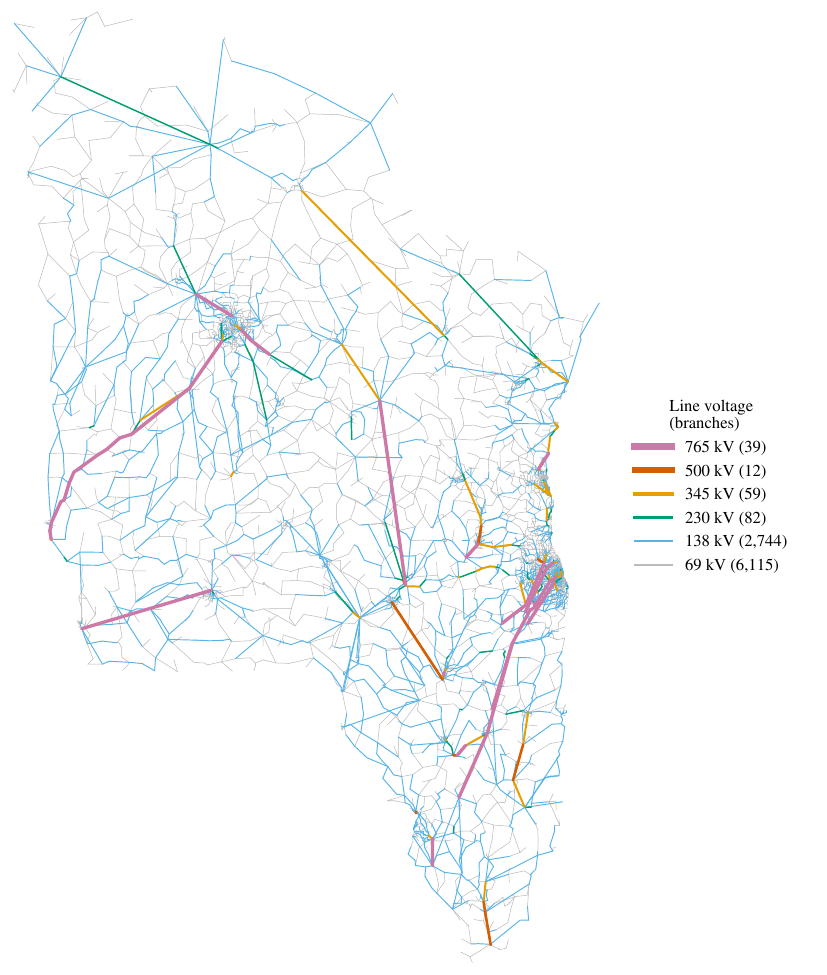}
  \caption{Geographic one-line of the MISO SCALE-Grid network ($6968$ buses,
  $9051$ branches), with lines drawn bus to bus and colored and weighted by
  voltage level; the legend gives the number of branches at each voltage.}
  \label{fig-MISO-OneLine}
\end{figure}

Figure~\ref{fig-MISO-OneLine} shows the geographic one-line of the MISO
SCALE-Grid network, drawn bus to bus and colored by voltage level, and
Supplementary Section~S5.4 shows five more footprints (AZNM, CAISO, NEISO, NWPP, and
NYISO). To visual inspection, the diagrams read as
plausible transmission systems, with a meshed high-voltage backbone overlaying a
denser lower-voltage network.

The FLO-Grid objective of Section~\ref{sec:flogrid-cost} penalizes only weighted
MW-miles of flow, with \emph{no} term for the fixed cost of acquiring a right-of-way
and building a new line, because that cost would reintroduce the binary variables
the method is designed to avoid. Even so, the constructed networks are sparse: each
linear-program solution adds a forest of lines (Section~\ref{sec:flogrid-inner}),
which suppresses most redundant construction.

However, two imperfections are visible. First, a few transmission lines are routed
in nearby parallel corridors, which the objective does not penalize; merging such
corridors into multi-circuit lines after construction could remove them. Second, the lower-voltage
networks keep some radial structure, most visibly at $69$~kV.
Both could be reduced by folding a subset of transmission-line $N$-1 contingencies
into the construction scenarios, so that the algorithm builds redundant paths only
where reliability requires them, at the cost of more computation
(Section~\ref{sec:conc}).

\subsection{Graph-metric validation: node degree}\label{sec:results-nodedegree}
\begin{figure}[t]
  \centering
  \includegraphics{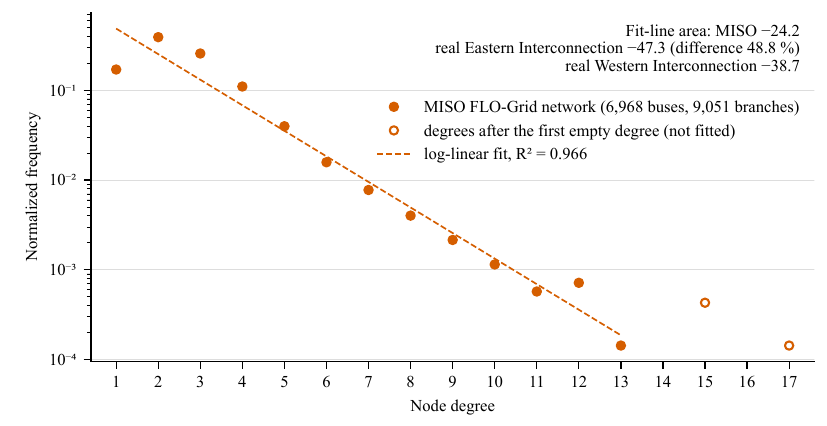}
  \caption{Normalized node-degree distribution of the MISO FLO-Grid network
  ($6968$ buses, $9051$ branches) on log-linear axes, with its log-linear fit
  ($R^2 = 0.966$). The fit-line area is $-24.2$ against $-47.3$ for the real
  Eastern Interconnection (a $48.8\%$ difference) and $-38.7$ for the real Western
  Interconnection. Open markers are degrees after the first empty degree, which
  the fit leaves out.}
  \label{fig-ND-MISO}
\end{figure}

\begin{figure}[t]
  \centering
  \includegraphics{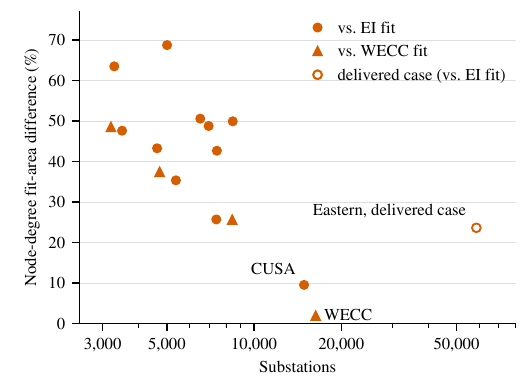}
  \caption{Node-degree fit-area difference against system size for the 15
  GridForge networks (13 regional footprints, CUSA, and WECC) and the delivered
  Eastern case (open marker; assembled from 10 regional networks by the same method). Circles are compared
  with the real Eastern Interconnection fit and triangles with the real Western
  Interconnection fit. Spearman $\rho = -0.546$ ($p = 0.035$, $n = 15$); with the
  Eastern case, $\rho = -0.618$ ($p = 0.011$, $n = 16$).}
  \label{fig-ND-Size}
\end{figure}

The node-degree distribution measures how densely transmission lines connect to
buses and is a standard structural descriptor for comparing synthetic and real
grids~\cite{Hines_2010_TopologicalElectricalStructure,Birchfield_2017_GridStructuralCharacteristics,Babaeinejad_2021_RealismMeasures}.
On log-linear axes, the normalized node-degree distribution of a transmission
network is close to a straight line, so an affine function is fit to each
distribution and the synthetic and real fits are compared
(Figs.~\ref{fig-ND-MISO} and~\ref{fig-ND-Size}). Supplementary Sections~S5.1
to~S5.3 list the size of each network, its node-degree values, and the fits of its
cycle-length distribution, which appear only there because a cycle basis is not
unique.

The synthetic distributions are strongly log-linear: across the 15 GridForge
networks (the 13 regional footprints and the combined CUSA and WECC systems), the
fit $R^2$ ranges from $0.92$ to $0.98$ (mean $0.96$), against $0.971$ and $0.981$
for the real Eastern and Western Interconnections. A high $R^2$ shows only that a
distribution is close to log-linear, a property many sparse spatial graphs share.
The fit-area difference, the percent difference between the areas under the
synthetic and real fits, measures how far the fitted line lies from the real one.
It averages $39.9\%$; AZNM, CAISO, NWPP, and WECC are compared with the Western fit
and the others with the Eastern fit. For MISO (Fig.~\ref{fig-ND-MISO}), the fit has
$R^2 = 0.966$ and a fit-line area (the signed area under the fitted line on the
log-linear axes) of $-24.2$ against $-47.3$ for the real Eastern
Interconnection, a $48.8\%$ difference.

Across the 15 networks, agreement improves with system size
(Fig.~\ref{fig-ND-Size}). The fit-area difference falls to $9.5\%$ for CUSA and
$1.9\%$ for WECC and decreases with the number of substations (Spearman
$\rho = -0.55$, $p = 0.035$), or $\rho = -0.62$ ($p = 0.011$) when the delivered
Eastern case of 58,583 substations is added. The fit $R^2$ shows no trend with
size ($\rho = 0.24$, $p = 0.38$). Two caveats apply. First, the networks are
compared with two different references. Second, the three largest systems (CUSA,
WECC, and the delivered Eastern case) combine regional runs, and their separated
voltage levels and transformers raise the branch-to-bus ratio of CUSA and WECC to
$1.715$ and $1.576$ (Table~\ref{tab:substations_by_region}), against $1.20$ to
$1.30$ for the regional footprints. Over the 13 regional footprints alone, the
trend is weak and not significant ($\rho = -0.30$, $p = 0.32$); most of the
correlation comes from CUSA and WECC. The delivered Eastern case was assembled in
2019 from 10 regional networks by the procedure of
Section~\ref{sec:scale-interconnect} and later edited in PowerWorld.

Note that only three large interconnections exist in the United States (the
Eastern and Western Interconnections and ERCOT), so the real population against
which any synthetic method can be measured is very small. The node-degree
agreement is therefore a measure of closeness to the real networks, not a formal
statistical hypothesis test.

\subsection{Electrical-performance validation: DC and DC optimal power
flow}\label{sec:results-electrical}
Structural realism does not guarantee electrical realism, so the networks are next
tested with a DC power flow and a DCOPF. The DCOPF minimizes total generation cost
subject to the DC-power-flow
balance of Section~\ref{sec:bg-relaxation}, generator limits, and branch thermal
limits, as in~\eqref{eq-DCOPF}, where $a_g,b_g,c_g$ are the quadratic cost
coefficients of generator $g$, $x_j$ is the series reactance of branch $j$, and $p_j^{\max}$ is its thermal limit.

\begin{subequations}\label{eq-DCOPF}
\begin{align}
\min_{P_G}\quad & \textstyle\sum_{g}\big(a_g + b_g P_{G,g} + c_g P_{G,g}^2\big)
   \label{eq-DCOPF-obj}\\
\text{s.t.}\quad & B_{\mathrm{Bus}}\,\theta = P_G - P_D, \label{eq-DCOPF-balance}\\
& P_{G,g}^{\min} \le P_{G,g} \le P_{G,g}^{\max},\quad \forall\, g,
   \label{eq-DCOPF-genlim}\\
& p_j = (\theta_i-\theta_k)/x_j,\quad \forall\, j=(i,k)\in E,
   \label{eq-DCOPF-flow}\\
& \lvert p_j\rvert \le p_j^{\max},\quad \forall\, j\in E. \label{eq-DCOPF-limit}
\end{align}
\end{subequations}

\begin{figure}[t]
  \centering
  \includegraphics[width=\linewidth]{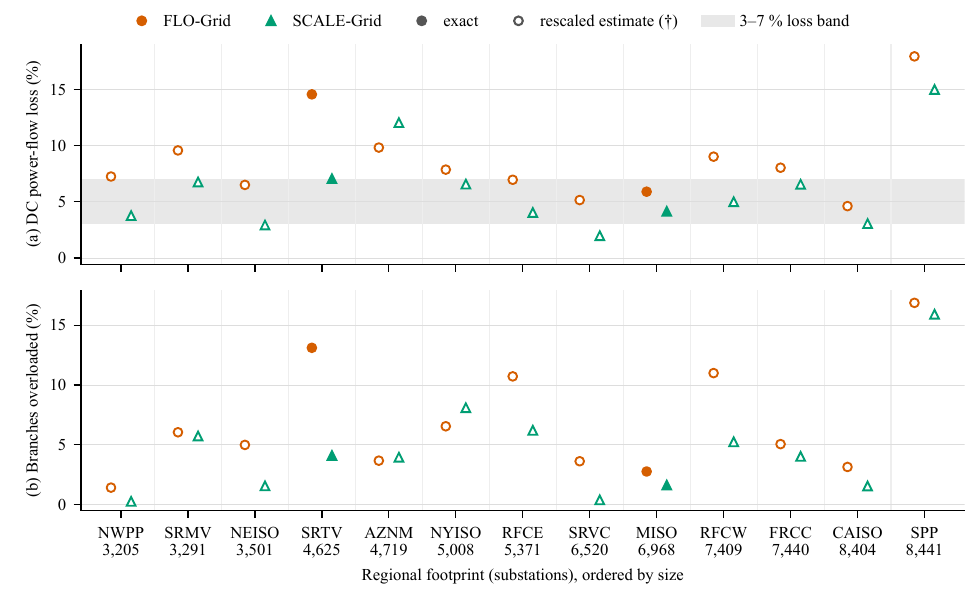}
  \caption{Mean DC power-flow loss (a) and share of overloaded branches (b) of the
  FLO-Grid and SCALE-Grid networks in each regional footprint, averaged over the
  footprint's evaluation scenarios. Footprints are ordered by number of
  substations, given under each name. Hollow markers are rescaled estimates
  ($^{\dagger}$); the MISO and SRTV values are exact. The shaded band marks $3\%$
  to $7\%$ loss, for orientation (Section~\ref{sec:results-electrical}).}
  \label{fig-DCPF-ByRegion}
\end{figure}

Figure~\ref{fig-DCPF-ByRegion} and Table~\ref{table-DCPF-Metrics-Summary} report
the fixed-injection DC power flow (Supplementary Section~S5.5 lists every
footprint), and Table~\ref{table-DCOPF-Metrics-Summary} the DCOPF. Published loss
figures give context, not a pass band: annual transmission losses in New York lie
between $1.7\%$ and $6.5\%$ of energy \cite{EPRI_2012_NYLosses}, and combined
transmission \emph{and} distribution losses averaged about $5\%$ in the United
States between 2018 and 2022 \cite{EIA_2023_TransmissionLosses} and reach about
$12\%$ for some utilities \cite{Wong_2011_TransmissionLosses}. These are annual
energy losses, whereas the scenarios here cluster around the annual peak and
minimum loads, and $I^2R$ losses grow roughly with the square of the load, so the
two are not directly comparable. Figure~\ref{fig-DCPF-ByRegion} shades $3\%$ to
$7\%$ for orientation only.

\begin{figure}[t]
  \centering
  \includegraphics[width=\linewidth]{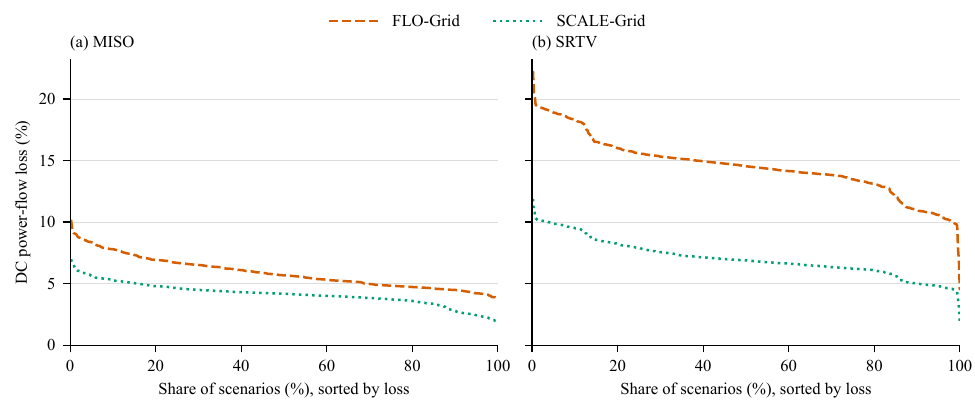}
  \caption{DC power-flow loss over the 486 evaluation scenarios, sorted from
  highest to lowest, for the FLO-Grid and SCALE-Grid networks of MISO (a) and
  SRTV (b).}
  \label{fig-DCPF-Duration}
\end{figure}

Sequential DC power-flow resizing (from
FLO-Grid to SCALE-Grid) lowers the loss in 12 of the 13 footprints and the share
of overloaded branches in 11. Across the 13 footprints, the mean loss falls from $8.70\%$ (FLO-Grid) to
$6.06\%$ (SCALE-Grid), and the mean number of overloaded branches falls from 536
to 363. On MISO, where the values are exact, the loss falls from $5.89\%$ to
$4.14\%$ and the overloaded branches from 249 to 146; on SRTV, the loss falls
from $14.55\%$ to $7.05\%$ and the overloaded branches from 788 to 246. Eight of
the 13 SCALE-Grid networks fall inside the $3\%$ to $7\%$ band (NEISO and SRVC
lie below it; AZNM, SPP, and SRTV lie above it), against five of the 13 FLO-Grid
networks.
However, resizing does not help everywhere (Supplementary Table~S7). It raises the
loss of AZNM from $9.82\%$ to $12.03\%$ and the number of overloaded branches in AZNM
(219 to 236) and NYISO (402 to 497). The overloads that remain are also larger in eight
of the 13 footprints, most in SRMV (mean overload $22.3\%$ to $60.6\%$) and RFCW
($47.2\%$ to $62.8\%$), since the resizing stops on the number of overloaded
lines, not on their size. SPP, the footprint with the most
substations and the highest load ($68.1$~GW), remains the worst, with $14.98\%$ loss
and $15.92\%$ of its branches overloaded. Figure~\ref{fig-DCPF-Duration} shows that the reduction
holds across operating points, not just on average. On MISO and SRTV, the
SCALE-Grid loss, sorted from highest to lowest, lies below the sorted FLO-Grid loss
over all 486 scenarios, by at least $1.10$ percentage points on MISO and $2.55$
points on SRTV.

\begin{figure}[t]
  \centering
  \includegraphics[width=\linewidth]{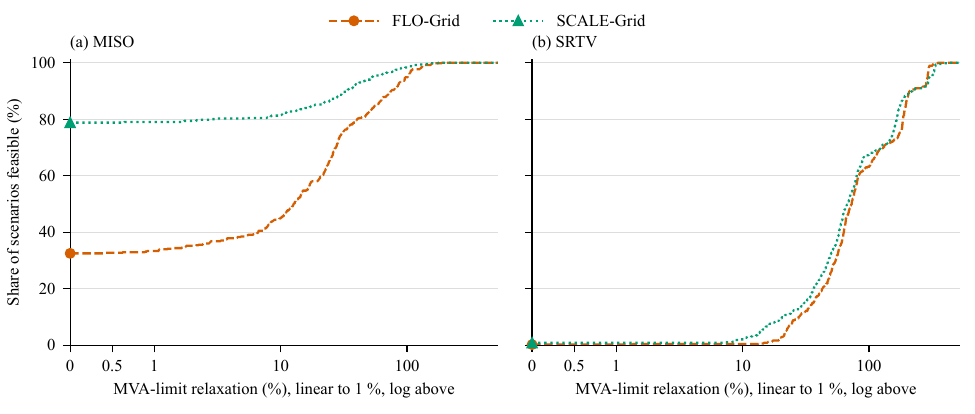}
  \caption{Share of scenarios that reach a feasible DCOPF within a given relaxation
  of branch MVA limits, for the FLO-Grid and SCALE-Grid networks of MISO (a) and
  SRTV (b), 486 scenarios each. The markers at zero relaxation give the share of
  scenarios feasible without relaxation.}
  \label{fig-DCOPF-Relax}
\end{figure}

When generators are instead permitted to redispatch through the DCOPF (the more
operationally meaningful test), Table~\ref{table-DCOPF-Metrics-Summary} reports
the two footprints with exact values, MISO and SRTV. When a scenario has no
feasible DCOPF at the stated limits, the MVA limits of all branches are scaled by
one common factor, found by the relaxation search of Supplementary
Section~S5.6. Where several dispatches reach the least cost, the one with the
least $I^2R$ loss is reported. This tie-break changes the mean loss by at most
$0.12$ percentage points, and a barrier and a dual-simplex solution agree
(Supplementary Section~S5.6). Resizing roughly halves the
DCOPF loss on both footprints, from $4.03\%$ (FLO-Grid) to $2.02\%$ (SCALE-Grid)
on MISO and from $5.54\%$ to $2.47\%$ on SRTV, both within the New York range of
$1.7\%$ to $6.5\%$, and it lowers the mean branch loading at the DCOPF solution
(Table~\ref{table-DCOPF-Metrics-Summary}). However, feasibility differs between the
two footprints. On MISO, the share of scenarios
that reach a feasible DCOPF without relaxing any MVA limit rises from $32.51\%$ to
$78.81\%$. On SRTV, it stays below $1\%$ ($0.21\%$ and $0.82\%$), and the
infeasible scenarios need their MVA limits relaxed by $106.68\%$ (FLO-Grid) and
$101.50\%$ (SCALE-Grid) on average. Figure~\ref{fig-DCOPF-Relax} shows the whole
distribution: on MISO, $81.28\%$ of the SCALE-Grid scenarios are feasible within a
$10\%$ relaxation, against $45.06\%$ for FLO-Grid, while on SRTV the largest
relaxation needed is $350.6\%$ (FLO-Grid) and $436.0\%$ (SCALE-Grid)
(Supplementary Section~S5.6). Closing this gap is future work
(Section~\ref{sec:conc}).

\begin{table}[t]
\centering
\caption{DC power-flow metrics of the FLO-Grid and SCALE-Grid networks: mean,
maximum, minimum, and standard deviation over the 13 regional footprints.}
\label{table-DCPF-Metrics-Summary}
\small
\begin{tabular}{@{}llrrrrrr@{}}
\toprule
 & \textbf{Network} & \textbf{\begin{tabular}[c]{@{}r@{}}Loss\\\%\end{tabular}}
 & \textbf{\begin{tabular}[c]{@{}r@{}}Branch\\\% loaded\end{tabular}}
 & \textbf{\begin{tabular}[c]{@{}r@{}}Over-\\loaded\end{tabular}}
 & \textbf{\begin{tabular}[c]{@{}r@{}}\% br.\\overld.\end{tabular}}
 & \textbf{\begin{tabular}[c]{@{}r@{}}Mean\\overld.\%\end{tabular}}
 & \textbf{\begin{tabular}[c]{@{}r@{}}Std\\overld.\%\end{tabular}} \\
\midrule
\multirow{2}{*}{Mean} & FLO-Grid    & 8.70\%$^{\dagger}$ & 29.3\%$^{\dagger}$ & 536$^{\dagger}$ & 6.84\%$^{\dagger}$ & 32.5\%$^{\dagger}$ & 31.7\%$^{\dagger}$ \\
                      & SCALE-Grid  & 6.06\%$^{\dagger}$ & 25.0\%$^{\dagger}$ & 363$^{\dagger}$ & 4.51\%$^{\dagger}$ & 37.5\%$^{\dagger}$ & 41.0\%$^{\dagger}$ \\
\midrule
\multirow{2}{*}{Max} & FLO-Grid    & 17.92\%$^{\dagger}$ & 34.9\%$^{\dagger}$ & 1,851$^{\dagger}$ & 16.87\%$^{\dagger}$ & 51.5\%$^{\dagger}$ & 54.2\%$^{\dagger}$ \\
                      & SCALE-Grid  & 14.98\%$^{\dagger}$ & 36.0\%$^{\dagger}$ & 1,747$^{\dagger}$ & 15.92\%$^{\dagger}$ & 62.8\%$^{\dagger}$ & 81.2\%$^{\dagger}$ \\
\midrule
\multirow{2}{*}{Min} & FLO-Grid    & 4.61\%$^{\dagger}$ & 25.0\%$^{\dagger}$ & 56$^{\dagger}$ & 1.41\%$^{\dagger}$ & 19.7\%$^{\dagger}$ & 19.7\%$^{\dagger}$ \\
                      & SCALE-Grid  & 1.98\%$^{\dagger}$ & 19.1\%$^{\dagger}$ & 10$^{\dagger}$ & 0.25\%$^{\dagger}$ & 8.7\%$^{\dagger}$ & 6.9\%$^{\dagger}$ \\
\midrule
\multirow{2}{*}{Std} & FLO-Grid    & 3.61\%$^{\dagger}$ & 3.3\%$^{\dagger}$ & 466$^{\dagger}$ & 4.47\%$^{\dagger}$ & 10.1\%$^{\dagger}$ & 9.9\%$^{\dagger}$ \\
                      & SCALE-Grid  & 3.58\%$^{\dagger}$ & 4.7\%$^{\dagger}$ & 431$^{\dagger}$ & 4.02\%$^{\dagger}$ & 16.4\%$^{\dagger}$ & 21.0\%$^{\dagger}$ \\
\bottomrule
\end{tabular}

\smallskip
{\footnotesize $^{\dagger}$Includes rescaled estimates (Section~\ref{sec:results-setup}).
The standard deviation of the overload is the mean over the scenarios of the
per-scenario standard deviation.\par}
\end{table}

\begin{table}[t]
\centering
\caption{DC optimal power-flow metrics of the FLO-Grid and SCALE-Grid networks on
MISO and SRTV, the footprints with exact values (486 scenarios each; where several
dispatches reach the least cost, the one with the least $I^2R$ loss is reported).
``\% MVA relax.'' is 100 times the recorded factor minus one, averaged over the
scenarios that need a relaxation to reach a feasible DCOPF; ``Std relax.'' is its
standard deviation over all 486 scenarios, counting the others as zero.}
\label{table-DCOPF-Metrics-Summary}
\small
\begin{tabular}{@{}llrrrrrr@{}}
\toprule
 & \textbf{Network} & \textbf{\begin{tabular}[c]{@{}r@{}}Load\\(GW)\end{tabular}}
 & \textbf{\begin{tabular}[c]{@{}r@{}}DCOPF\\loss \%\end{tabular}}
 & \textbf{\begin{tabular}[c]{@{}r@{}}\% cases\\feasible\end{tabular}}
 & \textbf{\begin{tabular}[c]{@{}r@{}}\% MVA\\relax.\end{tabular}}
 & \textbf{\begin{tabular}[c]{@{}r@{}}Std\\relax.\end{tabular}}
 & \textbf{\begin{tabular}[c]{@{}r@{}}Branch \%\\loaded\end{tabular}} \\
\midrule
\multirow{2}{*}{MISO} & FLO-Grid   & 36.02 & 4.03\% & 32.51\% & 36.54\% & 32.96\% & 22.57\% \\
                      & SCALE-Grid & 36.02 & 2.02\% & 78.81\% & 40.50\% & 22.76\% & 15.65\% \\
\midrule
\multirow{2}{*}{SRTV} & FLO-Grid   & 34.68 & 5.54\% & 0.21\% & 106.68\% & 81.13\% & 26.09\% \\
                      & SCALE-Grid & 34.68 & 2.47\% & 0.82\% & 101.50\% & 85.99\% & 17.55\% \\
\bottomrule
\end{tabular}
\end{table}

\subsection{Line-mileage sanity check}\label{sec:results-mileage}
As a final structural check, Table~\ref{table-metrics-busbranch-summary} compares
the voltage mix and mean line lengths of the GridForge networks with the real
Eastern and Western Interconnections. The mean line lengths are close to the real
ones (at 138~kV, $9.1$~mi against $7.3$~mi and $7.8$~mi; at 500~kV, $50.6$~mi against
$37.9$~mi and $49.2$~mi). The voltage mix is the main structural limit. The FLO-Grid
networks place $85.3\%$ of their lines at 69~kV, against $42.6\%$ in the Eastern
and $31.9\%$ in the Western Interconnection, and only $11.4\%$ at 138~kV, against
$44.6\%$ and $47.4\%$. After resizing, the $69$~kV share ranges from $36.2\%$ (SRTV)
to $80.3\%$ (CAISO), and $69$ and $138$~kV together hold $87.6\%$ (SRMV) to
$98.7\%$ (NEISO) of the lines (Supplementary Table~S13). The MISO network of
Fig.~\ref{fig-MISO-OneLine} has $67.6\%$ ($6115$ of $9051$ lines) at $69$~kV, at the
top of the band that steers the resizing (Section~\ref{sec:scale-sdcpf1}). Studies that depend on the
high-voltage backbone should take this mix into account. The bus shares follow from
assigning voltage to lines rather than substations and are expected to change when
a bus is created for each voltage level (Section~\ref{sec:scale-buses}), a stage not
evaluated here. Supplementary Section~S5.7 gives the values for every network.

\begin{table}[t]
\centering
\caption{Voltage mix and mean line length by voltage level (kV): mean of the 15
FLO-Grid networks (13 regional footprints, CUSA, and WECC) against the real
Eastern (EI) and Western (WECC) Interconnections. Shares are in percent of each
network's lines or buses; lengths are in miles. The real Western system has no
765~kV lines.}
\label{table-metrics-busbranch-summary}
\begin{tabular}{@{}llrrrrrr@{}}
\toprule
\textbf{Network} & \textbf{Metric} & \textbf{69} & \textbf{138} & \textbf{230} & \textbf{345} & \textbf{500} & \textbf{765} \\
\midrule
GridForge (mean of 15) & \% of lines & 85.3 & 11.4 & 2.1 & 1.2 & 0.5 & 0.1 \\
 & Mean length (mi) & 3.7 & 9.1 & 12.2 & 22.3 & 50.6 & 65.4 \\
 & \% of buses & 81.5 & 13.7 & 3.0 & 1.8 & 0.8 & 0.1 \\
\midrule
Real EI & \% of lines & 42.6 & 44.6 & 8.9 & 2.9 & 0.7 & 0.2 \\
 & Mean length (mi) & 4.7 & 7.3 & 13.5 & 29.1 & 37.9 & 88.9 \\
 & \% of buses & 53.2 & 36.7 & 7.2 & 2.1 & 0.5 & 0.1 \\
\midrule
Real WECC & \% of lines & 31.9 & 47.4 & 17.0 & 1.2 & 2.4 & 0.0 \\
 & Mean length (mi) & 6.0 & 7.8 & 19.9 & 62.2 & 49.2 & --- \\
 & \% of buses & 50.8 & 34.8 & 11.6 & 1.1 & 1.7 & 0.0 \\
\bottomrule
\end{tabular}
\end{table}

\subsection{Scalability}\label{sec:results-scaling}
\begin{figure}[t]
  \centering
  \includegraphics[width=\linewidth]{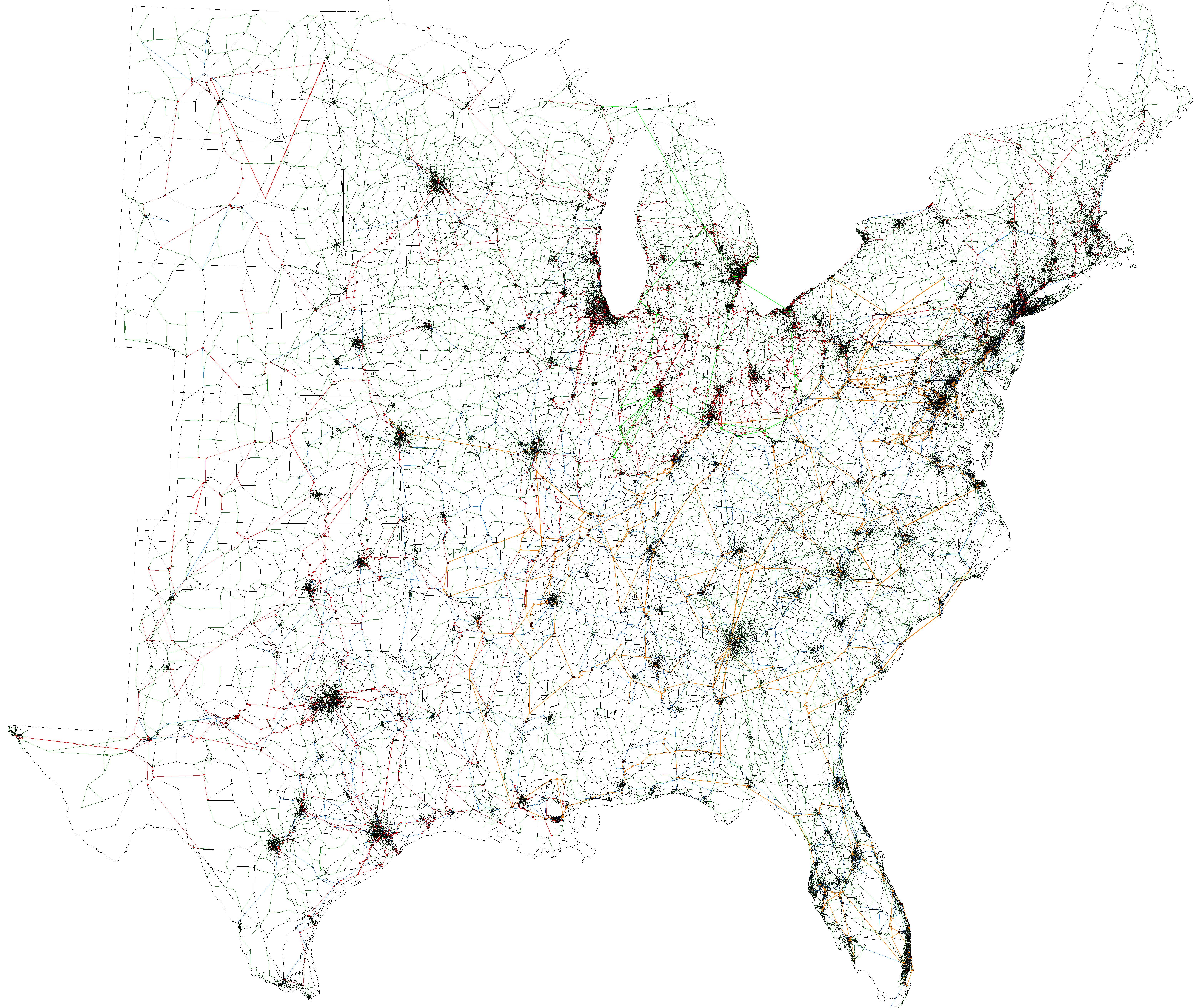}
  \caption{One-line diagram of the delivered $78{,}484$-bus synthetic Eastern United
  States case, the largest system built by the method, colored by voltage level
  (black: 69~kV; dark green: 138~kV; blue: 230~kV; red: 345~kV; orange: 500~kV;
  bright green: 765~kV; gray: state borders).}
  \label{fig-EI-OneLine}
\end{figure}

Each regional network was built on a single machine (Section~\ref{sec:results-setup}).
Table~\ref{table-Scalability} gives the two logged runs in which an earlier version of
FLO-Grid built a whole multi-region footprint from scratch, each in under a day on a
single 32-core server. These runs produced earlier variants, not the published
networks, which were assembled from regional runs
(Section~\ref{sec:scale-interconnect}), so no end-to-end build time exists for the
published networks. The first linear program of a footprint has two variables per
candidate and one equality row per substation, up to 40.3 million variables against
7409 rows in RFCW at a $230$~mi cap (Supplementary Table~S12). The 2019 builds did
not log the number of solves or the time of each stage. The networks range from 250
to 78{,}484 buses (Fig.~\ref{fig-EI-OneLine}).

\begin{table}[t]
\centering
\caption{Wall-clock time of the two logged complete FLO-Grid runs from scratch.
Both used an earlier FLO-Grid version and CPLEX 12.6.3.0 with 29 threads; their
outputs are earlier variants, not the published networks.}
\label{table-Scalability}
\small
\begin{tabular}{@{}lrrll@{}}
\toprule
\textbf{Footprint} & \textbf{Substations} & \textbf{Wall-clock time} & \textbf{Server} & \textbf{Run date}\\
\midrule
Western (WECC) & 16{,}328 & 19~h 21~min 35~s & 32 cores, 3.40~GHz, 256~GB & April 19, 2019\\
Midwest Valley (MVA) & about 15{,}000 & 19~h 27~min 42~s & 32 cores, 2.40~GHz, 256~GB & April 22, 2019\\
\bottomrule
\end{tabular}
\end{table}

\subsection{The open model library}\label{sec:results-epigrids}
The networks of this paper are listed in Supplementary Table~S11 with the stage
at which each was evaluated and where it is released. The delivered cases are in
the public repositories of the data availability statement, and the 13 regional
networks at both evaluated stages, with the scripts that compute every metric of
this section, are in a separate data deposit. They are part of a larger open
library, released without restriction on redistribution, whose complete release
and full multi-metric validation are future work (Section~\ref{sec:conc}). The
models contain no CEII, and their corridors correspond to no real line, so they may
be published alongside any study that uses them.

\section{Conclusion}\label{sec:conc}

The goal of this paper is to construct open, continental-scale synthetic power
grids with a method whose computational core is a linear program, since the exact
model of power-system construction is a mixed-integer, nonlinear, nonconvex program
that is intractable at scale (Section~\ref{sec:bg}). GridForge selects lines with Garver's
network-flow linear program, FLO-Grid, applied to every length-feasible bus pair of
a footprint (Section~\ref{sec:flogrid}), and then builds a fully parameterized,
multi-voltage network sized against DC power flow with the sequential DC power-flow
layer SCALE-Grid (Section~\ref{sec:scalegrid}).
The synthetic node-degree distributions are nearly as strongly log-linear as those
of the real Eastern and Western Interconnections, with linear-fit $R^2$ between
$0.92$ and $0.98$ against $0.971$ and $0.981$ (Section~\ref{sec:results}).
Sequential DC power-flow resizing lowers the mean DC power-flow loss across the 13
regional footprints from $8.70\%$ to $6.06\%$, although five of the 13 SCALE-Grid
networks lie outside the $3\%$ to $7\%$ range shaded in Fig.~\ref{fig-DCPF-ByRegion}.
It lowers the DC optimal power-flow loss on MISO from $4.03\%$ to $2.02\%$ and on
SRTV from $5.54\%$ to $2.47\%$, both within the $1.7\%$ to $6.5\%$ range of annual
transmission losses reported for New York. Models from $250$ to $78{,}484$ buses
were constructed on single machines, without a compute cluster, and are freely
redistributable.

Each solution of the inner linear program is optimal for the transportation
objective, but the multi-level scheme and the conductor and rating assignment of
SCALE-Grid are heuristic and carry no global-optimality guarantee, the price of
avoiding a nonconvex mixed-integer program. The most visible structural artifacts
are a small number of near-parallel corridors, some residual radial structure at
the lowest voltage levels, and occasional high-degree ``starburst'' buses at large
generators. They are limited in extent, given that the objective carries \emph{no}
integer decision variables and \emph{no} fixed-cost term for the capital expense of
a new line, and Section~\ref{sec:results-oneline} gives the remedies.

Five further limitations bound what Section~\ref{sec:results} shows. First, it
evaluates the FLO-Grid and first-round SCALE-Grid stages, not the delivered
multi-voltage cases, whose DC $N$-1 and DCOPF feasibility is a design goal rather
than a measured result; on SRTV, only $0.82\%$ of the scenarios reach a feasible
DCOPF without relaxing line limits. Second, no other synthetic family and no
mixed-integer solution is compared on the same footprints. Third, no network is
shown to solve an AC power flow. Fourth, the cost weights and three other targets
were calibrated (Section~\ref{sec:flogrid-cost}), so the DC power-flow and DCOPF
results are the independent tests, and $85.3\%$ of the FLO-Grid lines are at
$69$~kV. Fifth, the 2019 runs recorded no random seed and not
every per-region setting, so the evaluated networks are released as built rather
than regenerated from the code, and 11 of the 13 footprints carry rescaled rather
than exact metrics.

Future work starts with evaluating the delivered cases with DC power
flow, DCOPF, a DC $N$-1 screen, and AC power flow, and comparing them with another
synthetic family and with a mixed-integer solution on small systems. It also
includes the complete open library of regional and interconnection-scale models,
with a full multi-metric validation across every footprint. A successor
formulation, a successive quadratic program, models the DC power flow, and so
Kirchhoff's voltage law, directly in the inner loop and keeps the $L_1$
regularizer to enforce sparsity. Smaller refinements include creating and splitting
buses within substations, a voltage-level assignment that minimizes the number of
transformers, transmission-line and transformer $N$-1 contingencies in the
construction loop, and lines added after termination with the planning-sensitivity
method of~\cite{Birchfield_2020_PlanningSensitivities}.

GridForge does not displace the open synthetic grids in circulation today, built by
statistical matching, metric-matched construction, recombination, and
reconstruction; it adds an optimization-based family to that collection. Every
construction method leaves its mark on the grids it builds, so a model trained on
grids from several methods should have a better chance of learning the physics of
the power system rather than the habits of one method.

\printcredits

\section*{Declaration of competing interest}
The authors declare that they have no known competing financial interests or
personal relationships that could have appeared to influence the work reported
in this paper.

\section*{Funding}
This work was supported by the Advanced Research Projects Agency--Energy (ARPA-E)
of the U.S. Department of Energy under the GRID DATA program, award
DE-AR0000717.

\section*{Data availability}
The delivered synthetic cases (Supplementary Table~S11) are available under the
Creative Commons Attribution 4.0 International license (CC BY 4.0) at
\url{https://github.com/WISPO-POP/SyntheticElectricNetworkModels} and
\url{https://electricgrids.engr.tamu.edu/electric-grid-test-cases/}. The 13
regional networks at the FLO-Grid and SCALE-Grid stages evaluated in
Section~\ref{sec:results}, with the scripts that compute every metric, are
archived at \url{https://doi.org/10.5281/zenodo.24096575}
\cite{Snodgrass_2026_GridForgeData}, the networks under CC BY 4.0 and the scripts
under the MIT license.
The GridForge
construction code is available under the MIT license at
\url{https://github.com/Snodgrass-Research-Group/GridForge-Public}, and version 1.0.0
is archived at \url{https://doi.org/10.5281/zenodo.23826484}
\cite{Snodgrass_2026_GridForgeCode}. Users of either are asked to give credit by
citing this paper.

\section*{Declaration of generative AI and AI-assisted technologies in the manuscript preparation process}
During the preparation of this work the authors used Claude (Anthropic) to copy
edit the text. After using this tool, the authors reviewed and edited the content
as needed and take full responsibility for the content of the published article.

\bibliographystyle{model1-num-names}
\bibliography{references-ArXiv}

\end{document}